\documentclass[pre,floatfix,twocolumn,noshowpacs,,superscriptaddress]{revtex4}

\usepackage{amssymb,amsfonts,amsmath,mathrsfs}
\usepackage{epsfig,graphicx}
\usepackage{multirow,color}

\begin{document}

\title{The evolution of interdisciplinarity in physics research}

\author{Raj Kumar Pan}
\affiliation{%
Department of Biomedical Engineering and Computational Science, Aalto University School of Science, P.O.  Box 12200, FI-00076, Finland
}
\author{Sitabhra Sinha}
\affiliation{The Institute of Mathematical Sciences, CIT Campus, Taramani, Chennai 600113, India}
\author{Kimmo Kaski}
\affiliation{%
Department of Biomedical Engineering and Computational Science, Aalto University School of Science, P.O.  Box 12200, FI-00076, Finland
}
\author{Jari Saram\"aki}
\affiliation{%
Department of Biomedical Engineering and Computational Science, Aalto University School of Science, P.O.  Box 12200, FI-00076, Finland
}

\begin{abstract}
Science, being a social enterprise, is subject to fragmentation into
groups that focus on specialized areas or topics. Often new advances
occur through cross-fertilization of ideas between sub-fields that
otherwise have little overlap as they study dissimilar phenomena using
different techniques. Thus to explore the nature and dynamics of
scientific progress one needs to consider the large-scale organization
and interactions between different subject areas. Here, we study the
relationships between the sub-fields of Physics using the Physics and
Astronomy Classification Scheme (PACS) codes employed for
self-categorization of articles published over the past 25 years
(1985-2009). We observe a clear trend towards increasing interactions
between the different sub-fields. The network of sub-fields also
exhibits core-periphery organization, the nucleus being dominated by
Condensed Matter and General Physics. However, over time
Interdisciplinary Physics is steadily increasing its share in the
network core, reflecting a shift in the overall trend of Physics
research.
\end{abstract}

\maketitle

\section*{Introduction}
Scientific progress has been seen both as a succession of incremental
refinements as well as a succession of epochs with relatively slow
or little change that
are punctuated by periods of revolutionary transitions. In Popper's
view~\cite{popper}, science proceeds by gradually falsifying competing
candidate theories, whereas Kuhn~\cite{Kuhn62} argues that during episodes
of ``normal science'', scientists gradually improve their theories within
the current framework until enough unexplainable anomalies emerge to call for a major paradigm shift. Such shifts have occurred on many scales, from scientific revolutions with global reverberations 
to smaller breakthroughs within specific fields or sub-fields of science.
However, this view ignores the possibility of entirely new avenues of
research emerging from new connections that are forged between apparently
disjoint areas of science. Thus, new paradigms may be born not only because
of evidence that contradicts existing theories, but also because entirely
new questions and theoretical frameworks appear. For example, consider the
rise of systems biology, driven by technological advances in data
acquisition and their analysis through computer algorithms, or the
emergence of network science that merges aspects from physics, computer
science, and social sciences. 

In this paper, we focus on the dynamics and emergence of connections between the various subfields of physics, and
perform a longitudinal analysis of the evolution of physics from 1985
till 2009. Our results are based on a study of the
papers appearing in the Physical Review series of journals (Physical
Reviews A, B, C, D, E,  Physical Review Letters and Review of Modern
Physics) published by the
American Physical Society during this period, with their Physics and
Astronomy Classification Scheme (PACS) numbers indicating
the subfields of physics to which they belong. If a paper is listed
under two different PACS codes, the two corresponding sub-fields are
considered to be connected by the paper. In this manner
we construct a set of annual snapshots of the 
networks of sub-fields in physics that are connected through all
papers that have been published in each year,
and study the evolution of these networks at multiple structural scales. 
In this way, we can focus on the big picture of
the evolution of physics in terms of changes in the nature of 
connections between its subfields, instead of the microscopic level
that is considered by the widely studied collaboration or citation 
networks~\cite{Newman01a,Redner98,Pan12,Redner05}.

We show that the network of the subfields of physics is becoming
increasingly connected over time, both in terms of link density and the
numbers of papers joining different subfields. Despite gradual changes in
the network density, composition, and degrees of individual nodes, all key
statistical distributions display scaling, indicating stationarity in the
underlying micro-dynamics~\cite{Gautreau09}.  It is seen that a substantial
and increasing fraction of new links connects nodes that belong to
dissimilar branches of the PACS hierarchy, reflecting a trend where
inter-disciplinarity between the subfields of physics clearly increases.
By applying the $k$-shell decomposition technique, we show that the core of
physics has been dominated by Condensed Matter and General Physics for the
entire period under study, with Interdisciplinary Physics steadily
increasing its importance in the core.  It is seen that a substantial and
increasing fraction of new links connects nodes that belong to dissimilar
branches of the PACS hierarchy, reflecting a trend where
interdisciplinarity between the subfields of physics clearly increases.  By
applying the $k$-shell decomposition technique, we show that the core of
physics has been dominated by Condensed Matter and General Physics for the
entire period under study, with Interdisciplinary Physics steadily
increasing its importance in the core.

\section*{Results}
We have analyzed
all published articles in Physical Review (PR)
journals~\cite{APS} from 1985 till the end of 2009 which are
classified by their authors as belonging to certain specific
sub-fields using the corresponding PACS codes.
The PACS is an internationally
adopted, hierarchical subject classification system of the American Institute
of Physics (AIP) for categorizing publications in physics and astronomy
\cite{PACS}.
It is primarily divided into 10 top-level categories that represent broad research
areas. Each of these categories are then divided into smaller
domains representing more specific fields of physics, which may be further 
split into even more specific sub-fields. Thus, each of these PACS codes
represent a specific sub-field of physics.
(for a detailed description of the data, see Methods). 
For constructing the networks of the different sub-fields, we consider 
the PACS codes as nodes, a pair of which are linked if an article
is classified by both these codes.
In these networks, the degree $k$ of a node corresponds to
its number of links, \emph{i.e.}~number of other PACS codes it is
connected to, 
and its strength $s$ to the total number of articles published with
the
PACS code. The numbers of papers sharing two PACS codes are accounted
for with the weight $w$ of their link. In order to study the time evolution of this system, we create
yearly aggregated networks by considering all the articles published in a
given year (see Methods).

\subsection*{Network-level evolution of the system}
\begin{figure}[t!]
\begin{center}
  \includegraphics[width=1.00\linewidth]{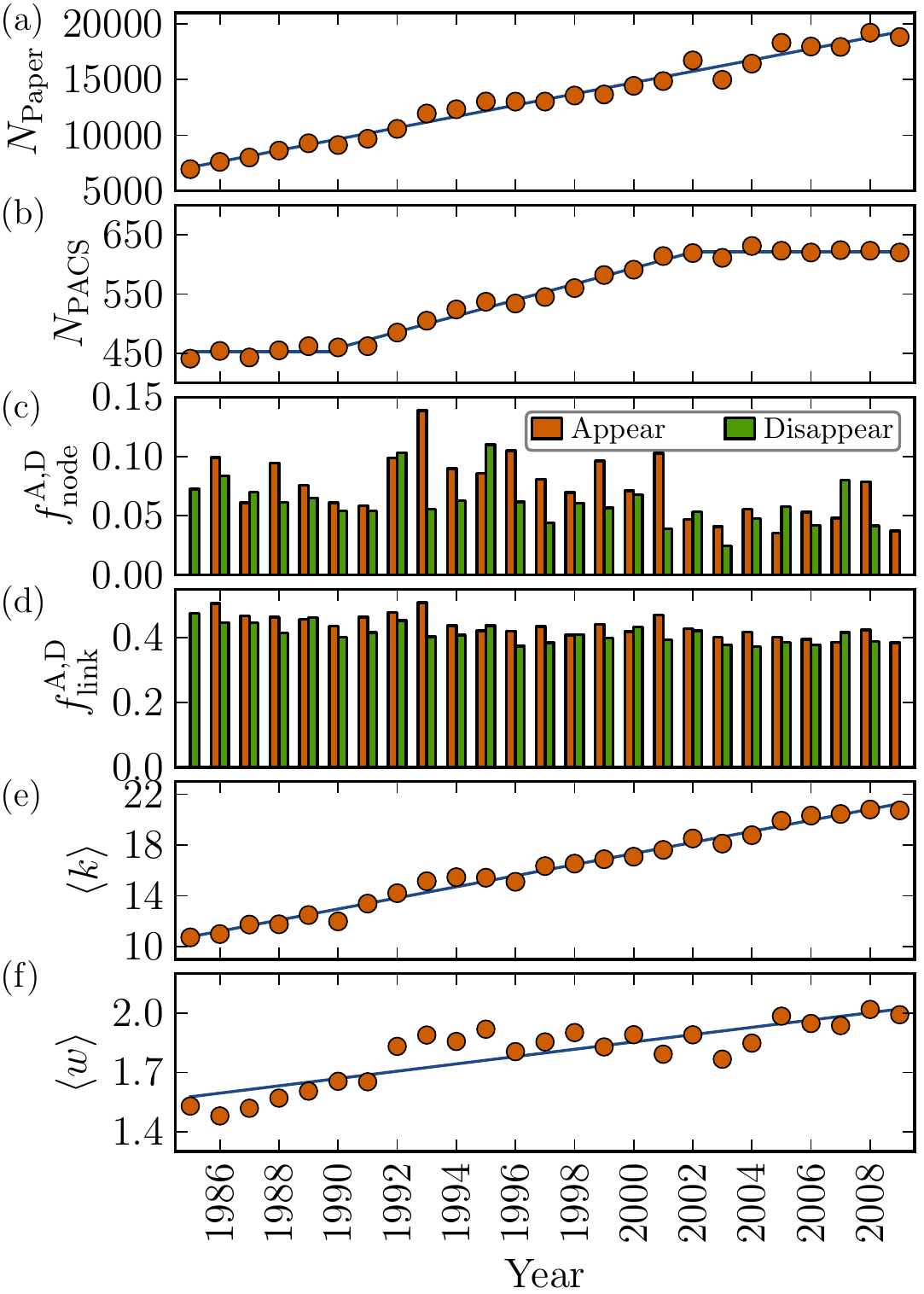}
\end{center}
\caption{The time evolution of various properties of the PACS network: (a) the number of published papers, (b) the number of PACS codes,
(c) the fraction of new and disappeared nodes, (d) the fraction of
new and disappeared links, (e) the average degree,
$\langle k \rangle$, and (f) the average link weight, $\langle w \rangle$.
The solid lines in (a), (e) and (f) denote a linear growth of  $\langle \Delta N_{\mathrm{Papers}}\rangle \approx 508$
papers per year, a yearly increase of $\langle \Delta k \rangle \approx 0.44$ of the average degree $\langle k \rangle$, and a yearly increase of $\langle \Delta w\rangle \approx 0.02$ of the average link weight $\langle w \rangle$, respectively. The solid line in (b) shows two roughly constant
regimes, interspersed by a period of average linear increase of $\Delta N_{\mathrm{PACS}}=13.5$
PACS codes per year between 1990-2002. 
Note that  $k$ and $w$ are heterogeneously distributed; see
Fig.~\ref{fig:degree_dist}.}

\label{fig:pacs_evolution}
\end{figure}

We begin by considering the evolution of the overall system properties
between 1985 and 2009. For these 25 years, the
total number of yearly publications $N_{\mathrm{Papers}}$ in all PR
journals has grown linearly~[Fig.~\ref{fig:pacs_evolution}(a)], while
the number of PACS codes $N_{\mathrm{PACS}}$ shows a linear
increase between 1990 and 2002, remaining roughly constant
before and after this period.
Note that this does not imply that the same codes have been
in use in all the years prior to 1990 or those after 2002, 
but rather that the number of new PACS codes that were introduced each
year were approximately balanced by the number of codes that were
discontinued that year. The
fraction of new and removed PACS codes each year is seen to fluctuate 
between 5\% and 15\% in Fig.~\ref{fig:pacs_evolution}(c). The yearly fractions
of new and disappearing links between PACS codes are higher,
fluctuating around $\sim 40\%$  [Fig.~\ref{fig:pacs_evolution}(d)]. 
When looking at network averages of the degree $\langle k \rangle$ and 
link weight $\langle w \rangle$ [Fig.~\ref{fig:pacs_evolution}(e),(f)], it
is seen that not only does the number of published papers grow, but
the network also becomes more connected, as both $\langle k \rangle$ and
$\langle w \rangle$ grow approximately linearly. As a consequence, the
average path length of the network decreases linearly (see Supplementary
Information). Thus, in general, the connectivity between different subfields of
physics is increasing with time. 

\begin{figure}[t!]
\begin{center}
\includegraphics[width=1.00\linewidth]{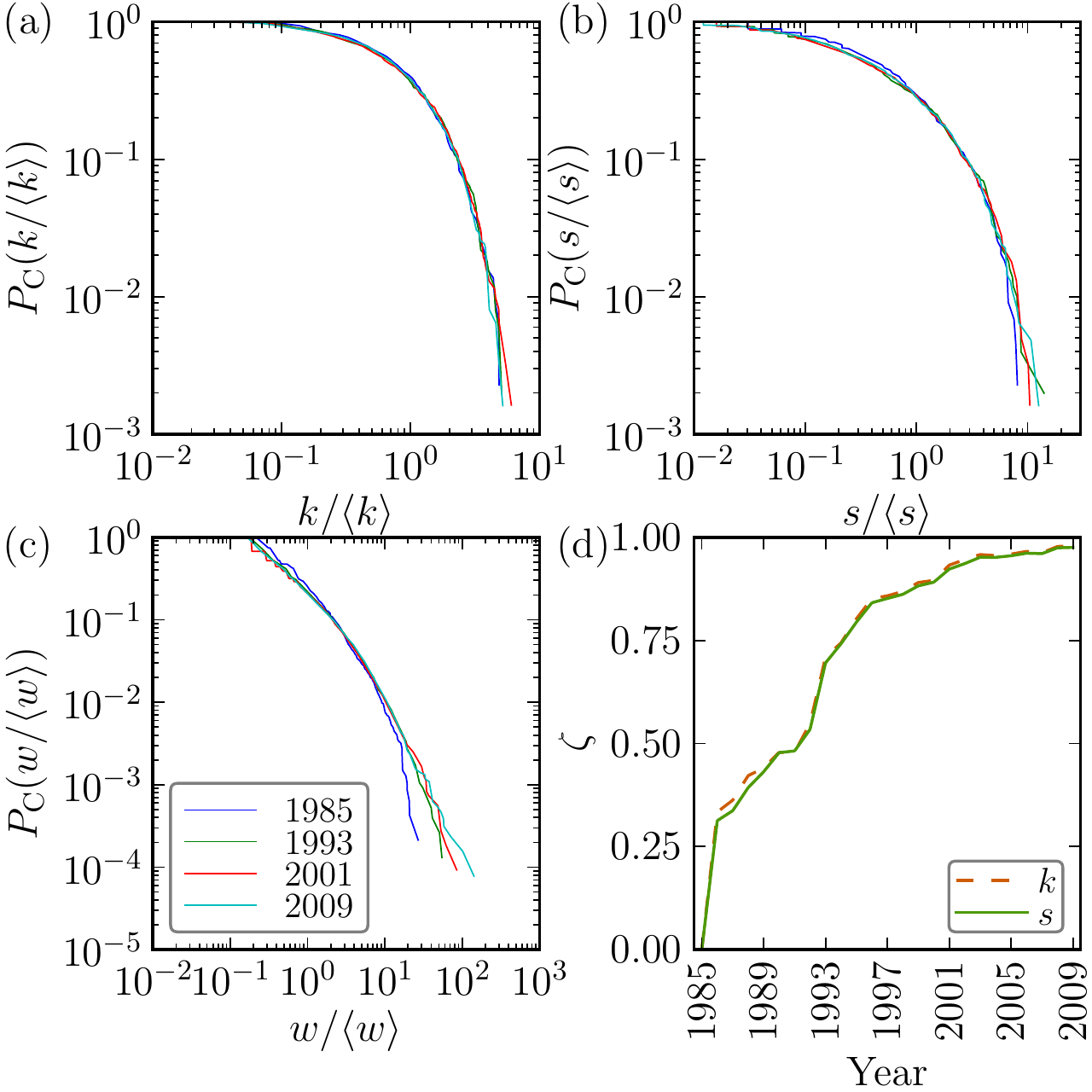}
\end{center}
\caption{Stationarity of the macro-level statistical distributions and variation at the micro level with time.
The cumulative distributions of (a) degree $k$,
(b) strength $s$ and (c) link weight $w$ of the PACS network, for four
different years. The curves have been scaled by their averages for the
given year.  (d) The dissimilarity coefficient $\zeta$ for the degree and
strength ranks of the nodes, between the year 1985 and subsequent years.} 
\label{fig:degree_dist}
\end{figure}

The scaled cumulative distributions of the key quantities (degree $k$, strength $s$, and
link weight $w$) are shown in 
Fig.~\ref{fig:degree_dist} for four different years. All distributions are broad and
indicate heterogeneity -- compared to the averages, some subfields of physics are much more connected
to the rest,  the links between some fields are stronger, and many more papers are published in some fields. 
Furthermore, the overlap of the rescaled distributions indicates that
although the averages of the distributions are growing over time, the
functional form of the distributions remains
similar~\cite{Gautreau09,Radicchi11}.
This is corroborated by comparing the Kolmogorov-Smirnov (KS) statistic of
the degree distribution of the yearly networks with each other and finding
that the KS distance stay at a low constant value~\cite{Massey51}.
A similar comparisons of the KS statistics of the strength distribution of
the yearly networks show similar behavior, although there is a slight
deviation from this general pattern for the year 1985 (see Supplementary
Information for further details).    
Hence,
although the composition of the system changes over time in terms of
nodes and links appearing and disappearing (Fig.~\ref{fig:pacs_evolution}), 
the functional shape of
the key distributions remain similar across the years, indicating
stationarity at the level of macro dynamics. 

In contrast to the relative invariance of the distributions, we
observe that over a long time-scale
the degrees and strengths of some nodes 
in the network 
increase or decrease in rank over time.
 Fig.~\ref{fig:degree_dist}(d) displays
the dissimilarity coefficient $\zeta$ of the degree ranks~\cite{Kossinets06} (see Methods) with 
respect to the year 1985 as a function of time; $\zeta \in [0,1]$ such that low values
indicate invariant node ranks.
It is seen that $\zeta$
increases monotonically with time, approaching $\zeta \approx 1$ towards the end. 
Thus, the degree ranks of the PACS codes change gradually over time
and become uncorrelated towards the end of the period under study,
indicating the presence of longer-term trends. Using the node strength to 
calculate $\zeta$ or calculating $\zeta$ between all pairs of years yields similar results (see Supplementary
Information).
We also compare the structural properties of the empirical
  PACS network with a randomized ensemble, in which PACS codes are reshuffled among
  papers. This is to see whether the observed properties of the
  network are expected to appear purely by chance as a consequence of
  the constraints inherent in the system.
  We found that in
  the randomized version there are many more links in the network
  compared to the empirical network leading to an increase in the clustering
  coefficient, decrease in the average link weight, and decrease in the average
  path length (see Supplementary Information).

\subsection*{Micro-level dynamics}

\begin{figure}[t!]
\begin{center}
\includegraphics[width=1.00\linewidth]{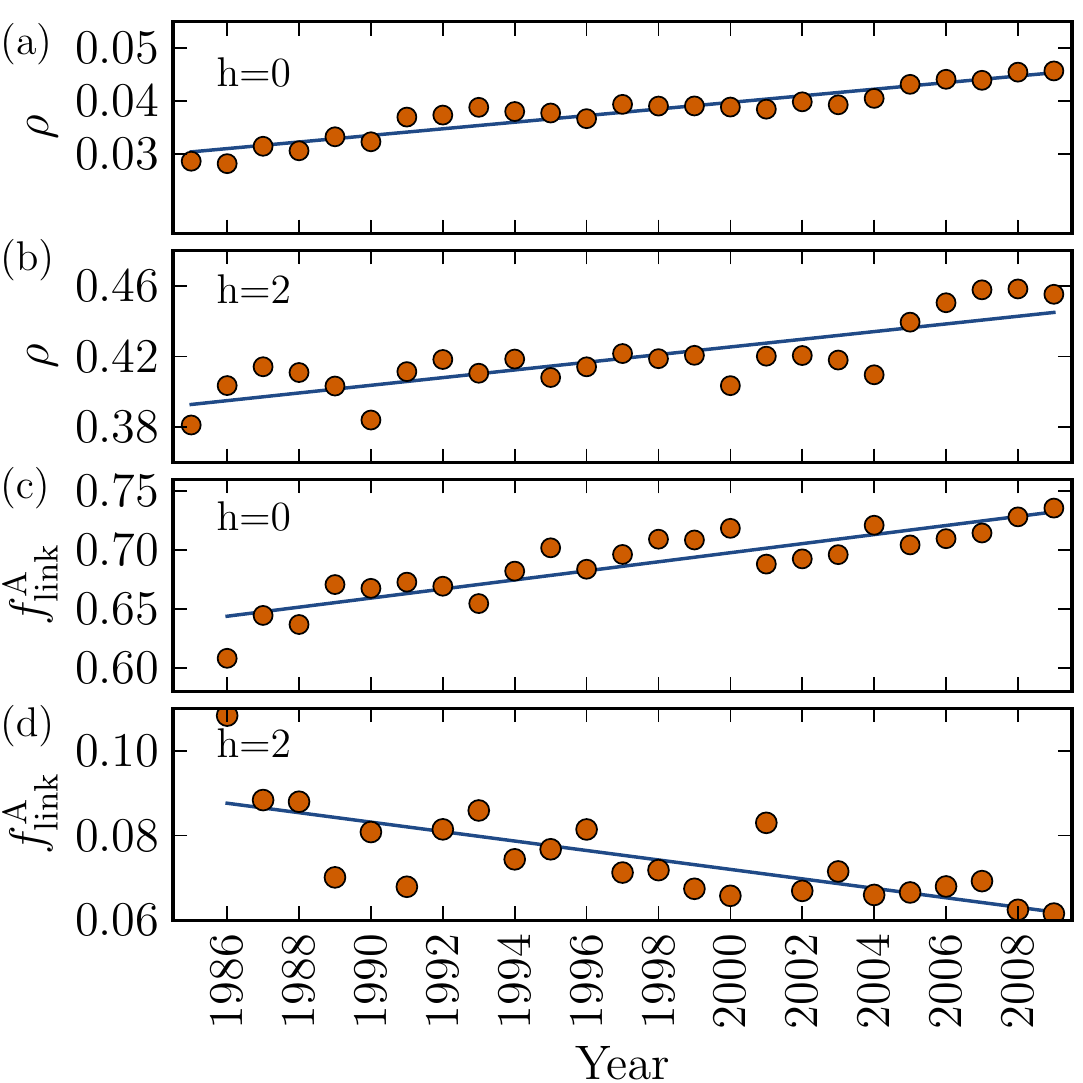}
\end{center}
\caption{The time evolution of network density and newly appearing links.
Evolution of the link density $\rho$ within the sets of nodes that are
hierarchically (a) dissimilar, and (b) similar up to second level.
Time dependence of the fraction of new links
$f_{\mathrm{Links}}^{\mathrm{A}}$ that connect nodes that are hierarchically
(c) dissimilar and (d) similar up to the second PACS level. The solid curves
indicate linear increase in (a), (b) and (c) with slope $6.2\times10^{-4}$,
$2.2\times10^{-3}$ and $3.9\times10^{-3}$, respectively and linear decrease
in (d) with slope $1.1\times10^{-3}$.} 
\label{fig:nLinksHie}
\end{figure}

\begin{figure*}[t!]
\begin{center}
  \raisebox{.75cm}{\includegraphics[width=0.40\linewidth]{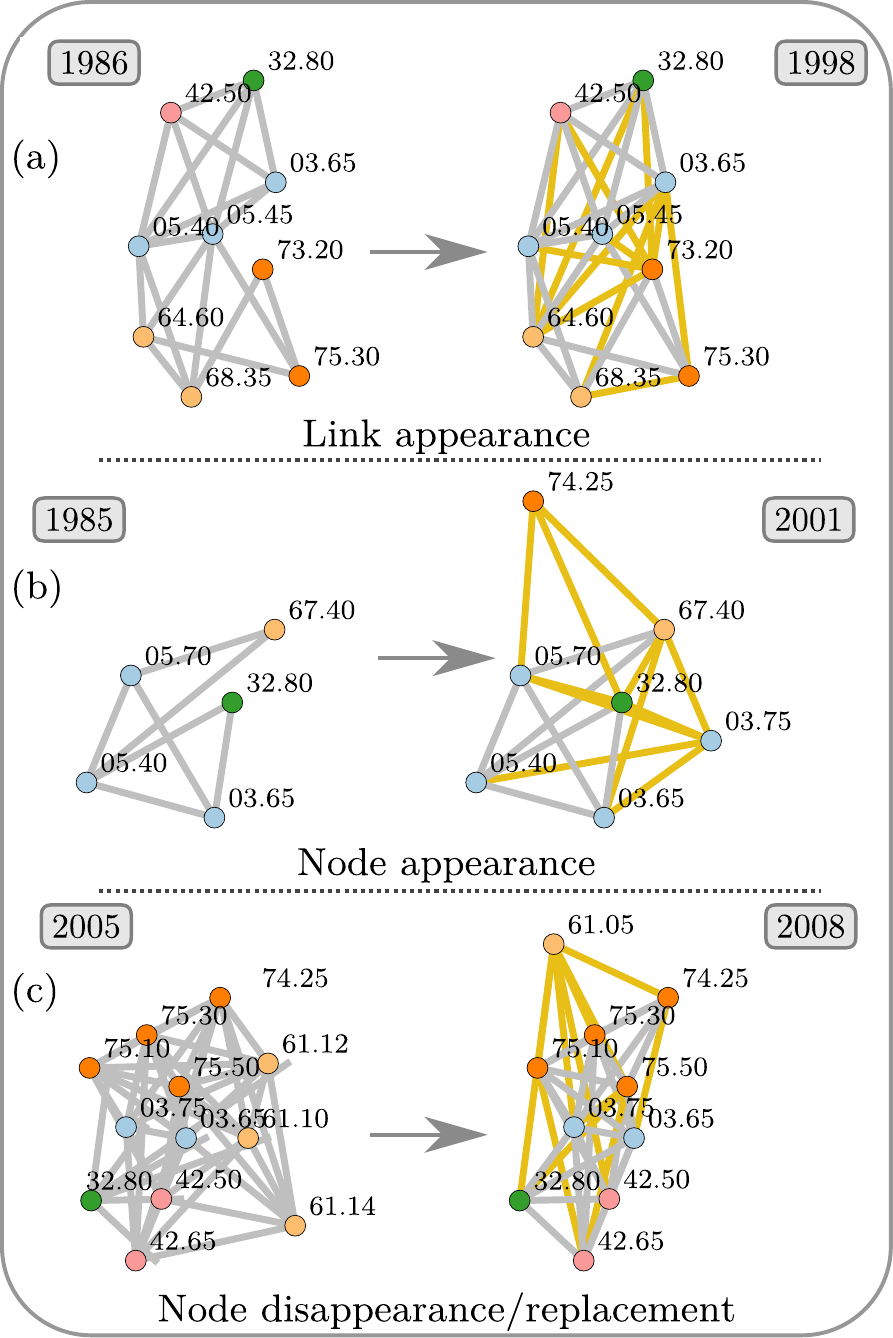}}
\includegraphics[width=0.50\linewidth]{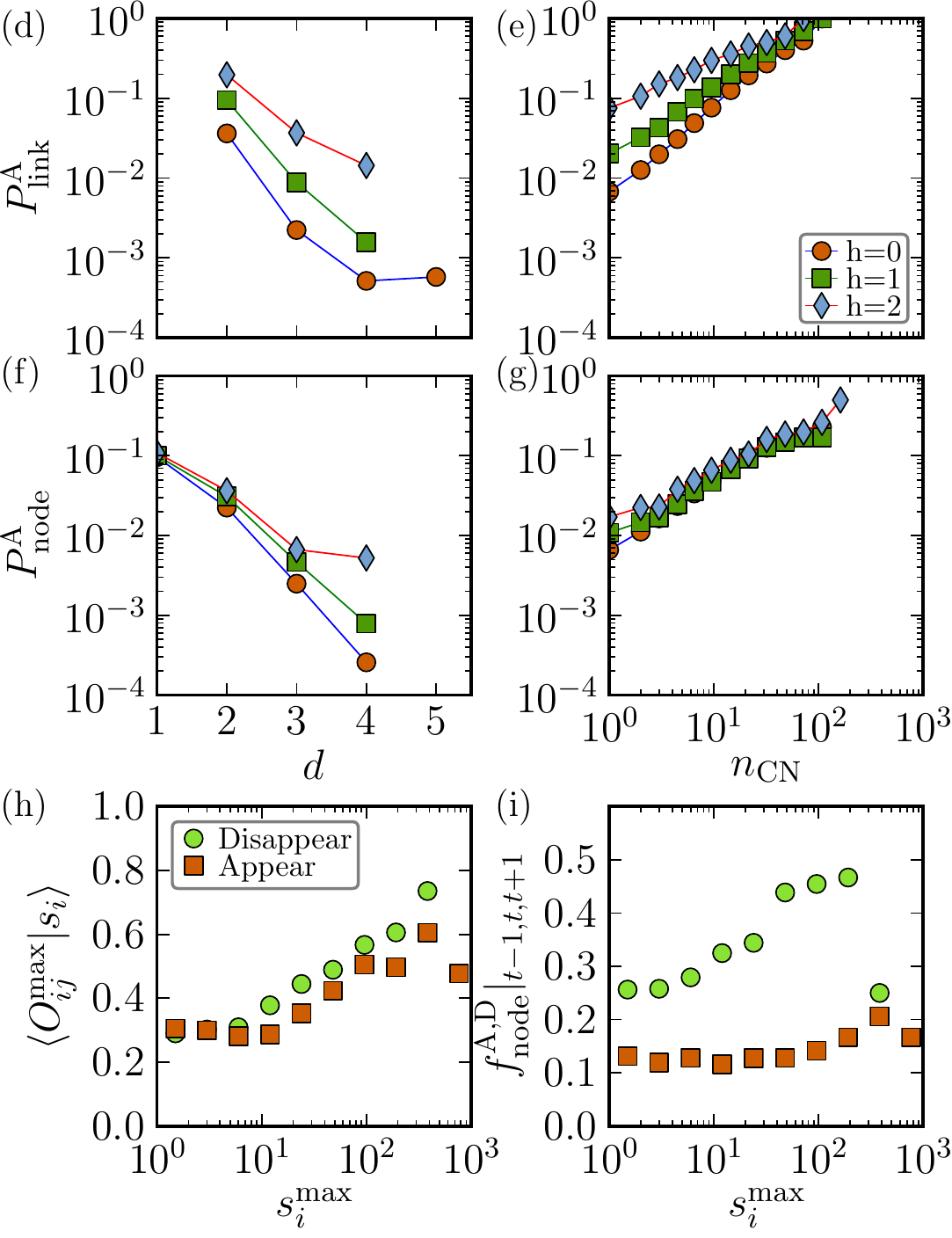}
\end{center}
\caption{Micro-dynamics in the PACS network. Examples of (a) the appearance of new
links that increase the density of a local neighborhood: a
large number of links appear between some sub-fields of condensed matter and general physics,
(b) the appearance
of new nodes (03.75, 74.25) and links increasing the density and (c) changes
in the network structure, where a new node (61.05) replaces several
existing nodes (61.10, 61.12, 61.14). 
Probability of link appearance as a function of the (d) distance $d$ and
(e) number of common neighbors $n_{CN}$ between the nodes.  Probability of
appearance of new nodes as a function of the (f) $d$ and (g) $n_{CN}$
between the nodes. The links categorized according to the
hierarchical similarity $h$ of the nodes they are connecting. 
(h) Similarity of discontinued nodes (circles) and newly introduced nodes (squares) with
their maximally similar counterpart nodes, as measured by the overlap
$O_{ij}^w$. The overlap is averaged over focal node strength. (i) The
fraction of maximally similar nodes that appear around the disappearance of
the focal node (circles) (from one year before the disappearance to one
year after) and the fraction of maximally similar nodes that have
disappeared around the time of appearance of the focal node (squares), again as a function
of focal node strength.}
\label{fig:examples}
\end{figure*}

Next we take a detailed look at the micro-dynamics of new and
disappearing links and nodes.
We take advantage of the hierarchical
nature of the PACS scheme (see Methods), and consider the hierarchical similarity $h$ of two
PACS nodes. Nodes are considered
dissimilar ($h=0$), if they belong to different main
branches of the PACS hierarchy and thus represent
very different subfields of physics. Nodes can also 
represent related subfields of physics and be
similar with respect to the first level of
hierarchy ($h=1$, i.e., they share their first PACS digit),
or similar with respect to the second level ($h=2$, i.e., they are even more
similar since they share
the first two PACS digits).
First, we focus on the link density $\rho$ of the network, defined for each similarity
class as the  number of links between nodes of the class normalized by the number
of pairs of nodes in the class. The evolution of the
 link density between dissimilar nodes ($h=0$) and nodes
belonging to the same second hierarchical level ($h=2$) is displayed in Figs.~\ref{fig:nLinksHie}(a)
and (b). For both cases, the density increases with time. As one would expect,
the link density for $h=2$ nodes
is far higher than that between dissimilar nodes. However, the relative increase of the 
density between the $h=0$ nodes is much higher, indicating an increasing trend
where new connections emerge between the main branches of physics. If the new links of each 
year are split into fractions according to whether they connect
similar or dissimilar sub-fields~[Fig.~\ref{fig:nLinksHie}(c-d)], it is seen that a substantial and increasing fraction of new links connects nodes
that belong to dissimilar branches of the PACS hierarchy ($h=0$), while the fraction of new links
joining similar PACS codes ($h=2$) decreases with time. Thus, there is an increase in interdisciplinarity between the subfields of
physics, as dissimilar branches of the PACS hierarchy are becoming
increasingly connected. This result holds even with 
a randomized null model that takes into account the different numbers of $h=0$ and $h=2$ nodes (see Supplementary Information).
Furthermore, this hierarchical connectivity and the increase in the
interdisciplinarity of the empirical network is lost in a randomized
network constructed by randomly shuffling the PACS codes across different
papers (see Supplementary Information).

Let us next address the role of network topology in the micro-dynamics. 
In particular, we want to see whether new links reflect the clustered
structure of the network, increasing the density of dense neighborhoods
as exemplified by the visualization of Fig.~\ref{fig:examples}(a). Additionally,
since the PACS numbers themselves evolve and new codes appear, local
clusters may also become increasingly connected if new nodes joining nearby
nodes appear, as in Fig.~\ref{fig:examples}(b).  The disappearance and appearance of nodes may also
reflect structural changes in the PACS system, such as code replacement
[Fig.~\ref{fig:examples}(c)]. 

First, we look for evidence for the mechanisms of Fig.~\ref{fig:examples}(a) and (b), where
new links are not randomly created, but follow a process where dense clusters of interlinked
PACS codes become even denser. For this,  we determine the geodesic distance $d$ (the number of links on the shortest path)
and the number of common neighbors $n_{\mathrm{CN}}$ for all pairs of nodes 
for each year, and count the number of pairs that are joined through a new link or through 
a new intermediate node in the following year.
This allows us to calculate the probabilities of link and connecting node appearance ($P_{\mathrm{links}}^{\mathrm{A}}$, $P_{\mathrm{node}}^{\mathrm{A}}$) aggregated over the data interval. Their dependence
 on the geodesic distance and number of common neighbors is 
shown in Fig.~\ref{fig:examples}~(d)-(g), where we have further divided all node pairs
into PACS similarity classes ($h=0,1,2$ as above). It is evident that the closer the nodes are and the more common neighbors
they have, the higher the likelihood of the appearance of a new direct link or a new joint neighbor
connecting the nodes. The mechanisms of Figs.~\ref{fig:examples}(a) and (b)
are thus common in the network, and new connections between the sub-domains of physics do not emerge in a random, 
uncorrelated fashion; rather, connectivity increases within clusters.
 Furthermore, the more similar a pair of nodes is with
respect to the PACS hierarchy, the higher the likelihood of new connections between them.
Similar features have also been seen in other networks, e.g., in social
networks new links are more likely to appear between nodes that are close,
that is, nodes that have common friends or share similar
interests~\cite{Kossinets06,Granovetter73,Liben-Nowell05}.

In order to study code replacement dynamics of Fig.~\ref{fig:examples}(c),
where discontinued codes are replaced by new codes that have a similar
connectivity pattern, we define a weighted version of the neighborhood
overlap $O_{ij}^w$ between a pair of nodes. This overlap is used to
determine the similarity in the neighborhood of two nodes so that
$O_{ij}^{w}=0$ if nodes $i$ and $j$ have no common neighbors, and
$O_{ij}^{w}=1$ if they have same set of common neighbors (see Methods). We
study all PACS codes that have been discontinued,
and first find their peak years $t^\ast$ with the highest number of papers.
For each PACS code $i$, we determine the network neighborhood $\Lambda_{i,t^\ast}$
corresponding to the peak year. We then calculate the overlap of this neighborhood
with the neighborhoods of all nodes in the network at year $t_i+1$, where
$t_i$ is the year when $i$ becomes discontinued. We then choose the node $j$
whose link pattern has the closest match with $i$ at its peak, as indicated
by the maximum overlap with $\Lambda_{i,t^\ast}$.  The average of this
maximum overlap $O_{ij}^{w\max}$ is displayed as a function of the strength
of the disappearing nodes $s_i$ in Fig.~\ref{fig:examples}(h).  The overlap
increases with the strength of the discontinued node.  Thus for
high-strength nodes, nodes of similar neighborhoods are present immediately
after their disappearance. These similar nodes are also usually introduced
around the time of discontinuation (see Fig.~\ref{fig:examples}(i)). Hence
high-strength PACS codes frequently get replaced rather than disappear
altogether; this can be taken indicative of gradual, continuous changes in
the subfields of physics. This might be due to the changing perceptions
about sub-fields as a result of gradually improving understanding of their
place in the general scheme of physics. These newly appearing codes have
connectivity similar to the disappearing PACS and also have many new
connections to other different sub-fields.

When a similar analysis is performed focusing on PACS codes that are newly
introduced, it is seen that nevertheless, the majority of new codes
correspond to emerging new subfields and do not appear to replace existing
codes (see Supplementary Information).



\begin{figure*}[t!]
\begin{center}
\includegraphics[width=1.00\linewidth]{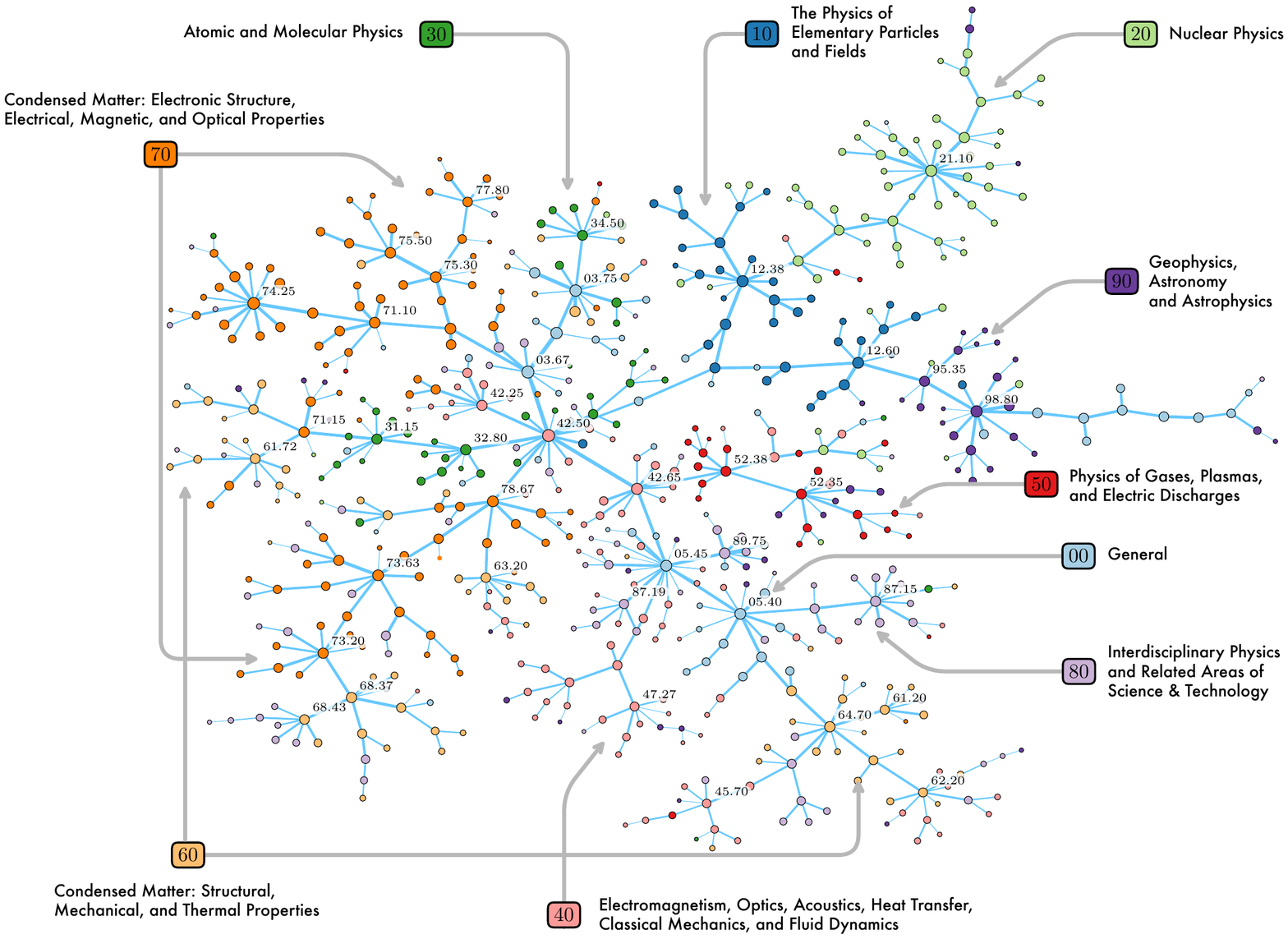}
\end{center}
\caption{The maximum spanning tree of the PACS network of 2009.}
\label{fig:mst}
\end{figure*}

\subsection*{Mesoscopic structure}

\subsubsection*{The Maximum Spanning Tree}

We now shift our focus from micro-dynamics towards the mesoscopic level and
begin by illustrating the structure of the PACS network with the help of
its maximum spanning tree (MST). The MST is a tree connecting all nodes of
the network while maximizing the sum of link weights; such trees can be
used to explore structural features in the data (see, e.g., \cite{OnnelaMST}). Figure~\ref{fig:mst}
displays the MST for the PACS network of the year 2009 (874 nodes). Some
structural features are apparent: first, as expected, PACS codes belonging
to the same broad categories are frequently connected in the MST; however,
there is mixing as well, especially in the central parts of the tree.
Second, the MST reflects the underlying cluster structure of the network.
There appears to be a branch that is well separated from the rest,
containing fields related to high-energy physics:
Physics of Elementary Particles and Fields, Nuclear Physics, and
Geophysics, Astronomy and Astrophysics. 
The rest of physics displays more mixing in the MST, the hub nodes 
being frequently related to General
Physics, Optics, and Condensed Matter.

\subsubsection*{$k^{s}$-shell analysis}

\begin{figure*}[t!]
\begin{center}
\includegraphics[width=0.70\linewidth]{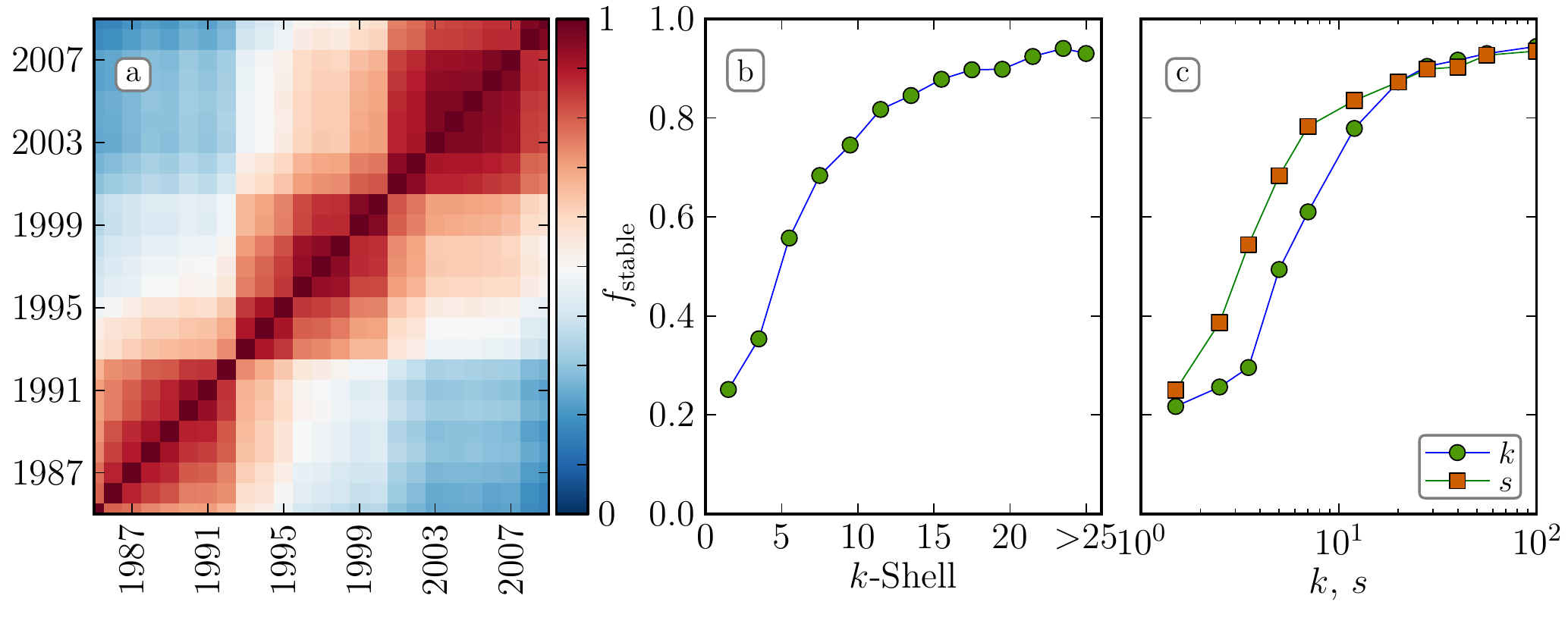}
\end{center}
\caption{(a) Matrix showing the correlation coefficient between the $k^s$-shell indices of the PACS codes for
different years. The fraction of the stable PACS codes that have remained present since their 
introduction as a function of its (b) $k^s$-shell index. (c) degree and
number of appearance.} 
\label{fig:kshellCorrelation}
\end{figure*}

\begin{figure*}[t!]
\begin{center}
\includegraphics[width=1.00\linewidth]{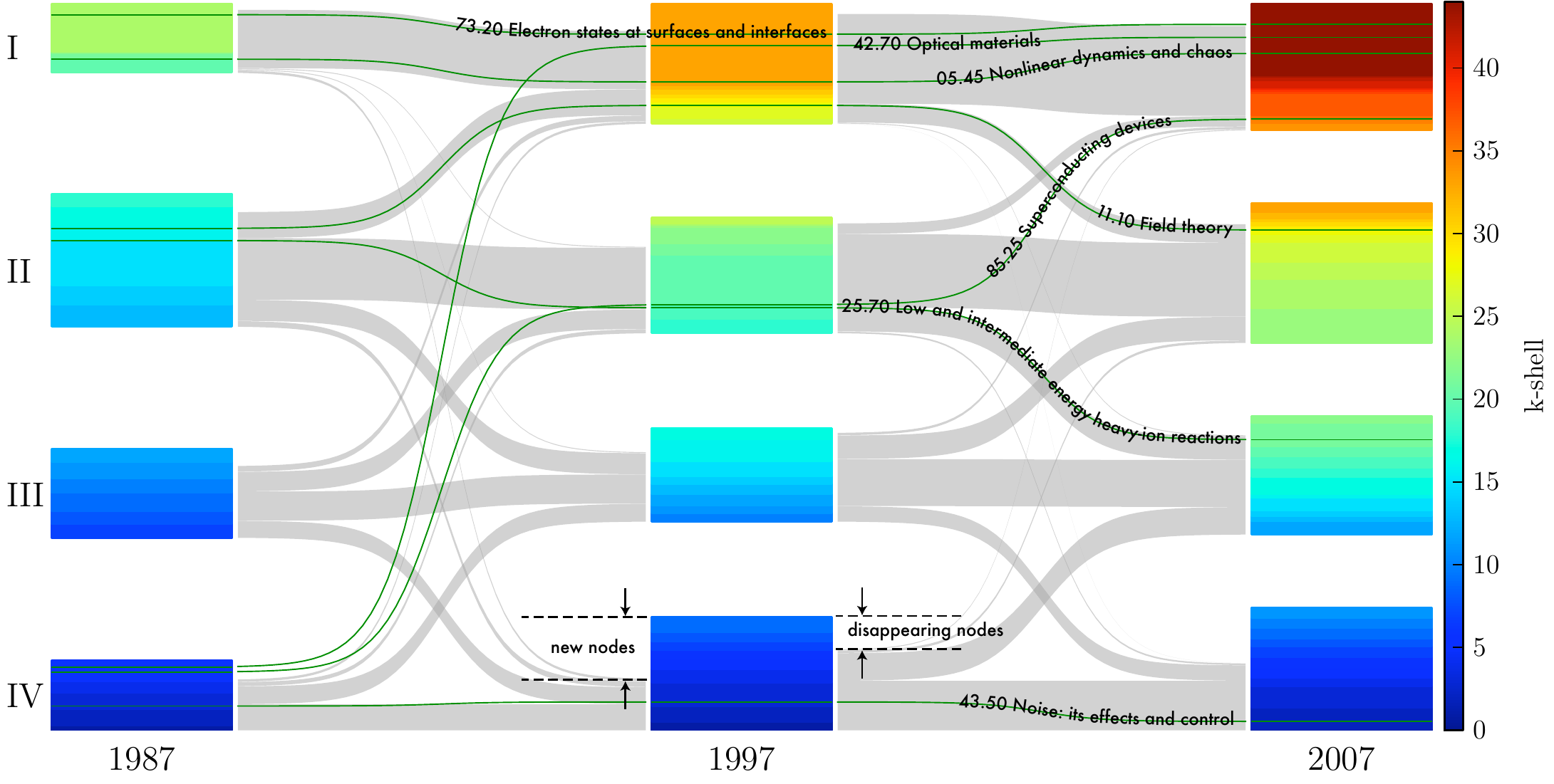}
\end{center}
\caption{Evolution of the $k^s$-shell indices of the PACS codes and 
the flows between $k^s$-shell regions between the years 1987, 1997, and 2007.
The PACS codes for
each year are divided into four different categories according to
their $k$-shell index (indicated by the color). The size of the block
indicates the number of codes in that category, and the widths of the 
shaded areas correspond to the fraction of migrating codes. The green lines
show $k^s$-shell trajectories for some specific PACS codes as examples.} 
\label{fig:kcoreEvolution}
\end{figure*}

Although the minimum spanning tree visualization of the network
provides some overview on the structural
organization of the relations between the different subfields of physics, it neither indicates the significance of
the nodes forming the core of the network nor gives us any information
regarding the temporal evolution of the structure.  For a better and more
detailed understanding, we perform $k$-core analysis~\cite{Bollobas,Seidman,Carmi07, KitsakNP2010} of
the evolving PACS network by decomposing the network for each year
into its $k^s$-shells (see Methods), such that a high $k^s$-shell index of a node
reflects a central position in the core of the network.

First, we want to establish that the $k^s$-shell indices of the PACS codes are
relatively stable over time and are thus suitable for analysis. 
To do this
we determine the correlation coefficients
between the $k^s$-shell indices of all the PACS codes and between different
years. In Fig.~\ref{fig:kshellCorrelation}~(a) the correlation coefficient
between different pairs of years are represented in terms of a matrix 
with the color of each cell
representing the corresponding correlation value.
The coefficient has a high value for neighboring years, so that
changes in the shell indices of nodes appear gradual over time 
rather than randomly.
Thus, the nodes having high or low $k^s$-shell index for year $t$ are more likely to
retain their 
index for the subsequent year $t+1$. 
Furthermore, the correlation matrix shows a block diagonal structure, 
indicating higher correlations for three periods, 1985-1992, 1993-2000 and
2001-2009. For analysis of $k^s$-shell regions (see below), we pick
one network corresponding to each of these periods.
The $k^s$-shell indices of PACS codes are also related to their
stability. We define a node as stable if it has been in use each year after its
introduction. Fig.~\ref{fig:kshellCorrelation}~(b) shows the fraction of stable
nodes calculated over the entire period 1985-2009 as a function of the $k^s$-shell index;
it is evident that the higher the order of the $k^s$-shell (and thus, the
closer it is to the nucleus of the network), the larger is the fraction of
stable nodes.
Note that, as the $k^s$-shell index of a node is related to its degree and
strength, nodes that have high degree or strength are also less likely to
get deleted and are more stable.

For studying the time evolution of the $k^s$-shells, we use the alluvial diagram method~\cite{Rosvall10}.
We divide the PACS codes into four categories based on their $k^s$-shell
indices by dividing the range of $k^s$ values into four groups of
approximately equal sizes. Thus
Region I  contains codes that are in the core of the
network ($k^s\in[\frac{3}{4}k_{\max}^s,k_{\max}^s]$), and
Regions II, III, and IV contain nodes with increasingly lower $k^s$-shell indices. 
The colored blocks of the alluvial diagram in
Figure~\ref{fig:kcoreEvolution} show the different regions for
three different years, with the size of each block representing the number of
PACS codes in the respective region. The sizes are increasing with time,
indicating an increase in the number of PACS codes. Furthermore, the maximum
shell index $k^s_{\mathrm{max}}$ has increased with time, as indicated by the
color of the $k^s$-shell indices for different years.

The shaded areas joining the $k^s$-shell regions represent flows
of PACS codes between the regions, such that the width of the flow
corresponds to the fraction of nodes. The total width of incoming
flow is less than the width of the corresponding region, because the
rest is made up by new PACS codes entering the network. Likewise, the gap between the
width of the block and total outgoing flow corresponds to discontinued
PACS codes. Here, it is seen that the core of the network, Region I, is remarkably
stable compared to the peripheral Region IV that displays a high
turnover of codes. Nodes that are in the core of the network are highly likely
to remain so, whereas peripheral nodes frequently either disappear or
migrate towards the core. Furthermore, a high fraction of new nodes first
appear in the peripheral region.

\begin{figure*}[ht!]
\begin{center}
\includegraphics[width=1.00\linewidth]{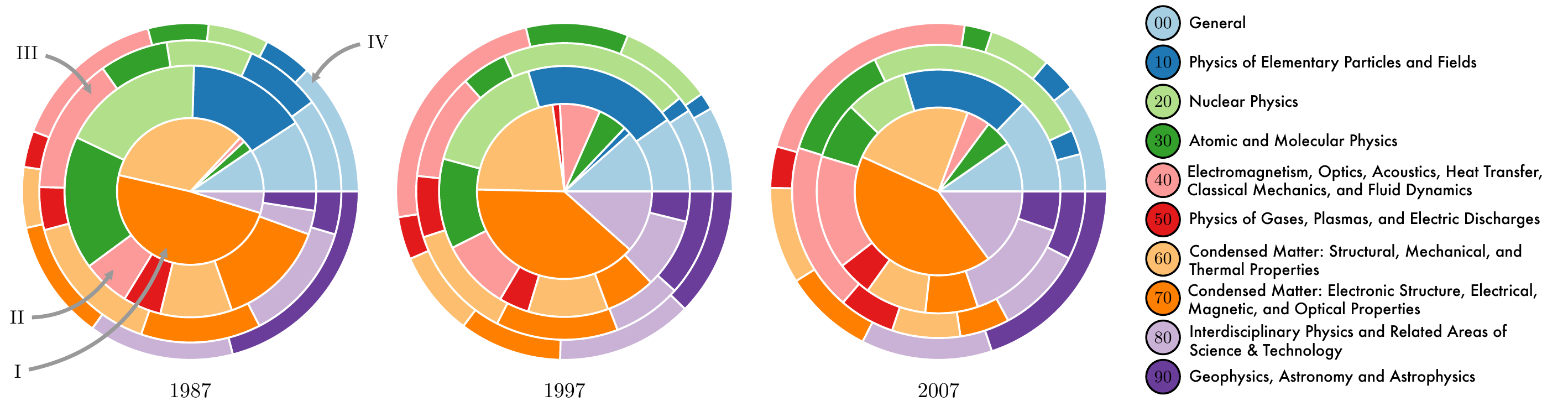}
\end{center}
\caption{Multi-level pie chart for year 1987,1997 and 2007 showing 
the composition of each of the PACS $k^s$-shell regions (I-IV),
such that the colors represent the first level of the PACS hierarchy.} 
\label{fig:kshellEvolution}
\end{figure*}

Next, we consider how the different branches of physics are positioned
with respect to the core-periphery organization of the PACS network 
and how their position has changed over time. 
Figure~\ref{fig:kshellEvolution} displays  multi-level
pie charts for three different years, where each level of the chart represents one of the $k^s$-shell regions
as above. The innermost layer
represents Region I, followed by Region II, Region III,  and
finally the outermost layer represents the peripheral Region IV. 
For each layer, we show the fraction of level-3 PACS codes belonging
to the different branches of physics as indicated by their first hierarchical
PACS level. 

The pie chart for the year 1987
shows that the core region I consists mostly of General Physics and 
Condensed Matter (PACS categories 00, 60 and 70), with 
a small contribution from categories 30 (Atomic and Molecular Physics), 40 (Electromagnetism etc), and 80 (Interdisciplinary Physics). In all other
regions, all branches of physics are present. 
For the network structure of 1997, we see that the contributions
of PACS categories 30, 40, and 80 have increased in the core region. 
Looking at the pie chart for the year 2007, we see that Interdisciplinary
Physics (80) has taken over an even larger fraction of the core.
The three main groups in the core are the two Condensed Matter categories
(60, 70) and Interdisciplinary Physics (80). At the same time, it is seen that Nuclear Physics (20)
has been moving towards the periphery, mainly contributing to Region III; this
is in line with its position in the MST of Fig.~\ref{fig:mst}. 
Thus, between 1987 and 2009, we see that 
Condensed Matter and General Physics have retained their position in the
very core of physics, while Interdisciplinary Physics
has been steadily moving towards the core, and Nuclear Physics has migrated
towards the periphery. Furthermore, Physics of Elementary Particles and
Fields (10) and Astrophysics (90) have retained their relative core position during
this period. Note that if the above pie charts are calculated on the 
basis of the total number of papers for each PACS code (see Supplementary
Information), no clear
evolution can be observed, as the codes are more homogeneously
distributed in the regions. This indicates that within each
hierarchical level-1 category,
there are level 3 PACS codes with highly varying volumes of publication
activity and this volume does not directly correspond to the position
of the code in the network.

\section*{Discussion}

We have studied the evolution of physics research in terms of interconnections
between its subfields from 1985 to 2009. 
We have shown that 
for yearly networks constructed from PACS codes,
although there are apparent dynamical changes in
the network, the key statistical distributions display remarkable stationarity.
The average number of links per code and average link weight show a steady increase, indicating increased
connectivity between different subfields of physics. 
In particular, the rate of
link formation between subfields that are distant in the PACS hierarchy
has increased, pointing out a clear trend of increased interdisciplinarity
within physics where its different branches are becoming increasingly interlinked.
This evolution does not appear random or uncorrelated; rather, within the branches
there are subfields that are joined together in clusters, and there is a tendency where
subfields in such clusters get connected through new links or new intermediate 
subfields with a high rate. 
The ``mesoscopic" or intermediate-scale analysis of the network
suggests an evolution towards increasing
interdisciplinarity in physics, and a detailed study of the properties of such growing
clusters would likely provide important insights into the evolution of physics.


At the mesoscopic level of the network, $k$-shell decomposition analysis
reveals some large-scale trends within physics discipline: the nodes
participating to the core of the network display the highest probability of
survival, whereas the peripheral region displays the largest turnover
associated with the discontinuations of older PACS codes and the appearance
of new ones, as well as, their migration towards the core. The nodes that
are in the core have a large number of connections to a large number of
other nodes, and thus a high $k$-shell index can be taken as indicative of
the importance of a PACS code compared to the ``rest" of physics. With this
interpretation it is natural that such high-$k$-shell subfields of high
importance are also subfields of high stability. In our data, the core of
the network has been dominated by those PACS codes that belong to the main
branches of Condensed Matter and General Physics for the entire period
under study.  However, we also note that there is an important trend of the
PACS codes belonging to Interdisciplinary Physics to steadily migrate
towards the core, so that at present these already occupy a significant
fraction of the core.

In conclusion, there has been an increase in the interdisciplinarity within
physics, as indicated by the evolution of interconnections between
different branches of physics. In addition there is an increase in the
importance of Interdisciplinary Physics that also has connections to fields
outside physics, as indicated by its share of the core in the PACS network.
Although it may be easy to identify candidate drivers for this evolution,
like the availability of vast amounts of digital data in several areas
(e.g., financial markets, social systems) and an increasing number of
problems requiring specialists from several fields within and outside
physics (e.g., problems related to energy, climate, and biophysics),
assessing their importance is beyond the scope of this study. It would be
especially interesting to see how the availability of research grants in
different sub-fields of physics correlate with our observations, and
whether the evolution of physics follows the amount of funding available
for its sub-areas or vice versa. This would require data about science
funding collated from many sources. In addition, the PACS codes represent
only one possible way to define the subfields of physics. Furthermore,
there may be delays between developments in physics and respective changes
in the PACS hierarchy. Nevertheless, we feel that it would be very
interesting to compare our results with a study of the network of
inter-relations between physics sub-fields constructed by using some other
data than the PACS codes and recent methods such as community structure
analysis of citation or co-authorship networks used to define the
subfields.

\section*{Methods}

\noindent
{\bf Data description:}
A PACS
code contains three elements: a pair of two-digit numbers separated by
``.'' and followed by two
characters that may be lower- or upper-case letters or ``$+$'' or ``$-$" signs.
The first digit of the first two-digit number denotes the main
category out of the 10 broad categories specified at the first level 
 and
the second digit gives the more specific field within that category. 
The second two-digit number specifies a narrower category
within the field given by the first two digits. The last two characters
may specify even more detailed categories up to the fifth level of hierarchy.
As an example, in the PACS code 05.45.-a, the first digit ``0" indicates ``General'', adding
the second digit ``05'', denotes ``Statistical physics, thermodynamics, and nonlinear dynamical
systems'' and 05.45.-a indicates ``Nonlinear dynamics and chaos'';
the ``-'' sign denotes the presence of one more level of hierarchy. 
Our source data comes in the form of the PACS codes of all published 
articles in Physical Review (PR)
journals~\cite{APS} of the American Physical Society from 1985 till the end
of 2009. 
In this study we use the PACS codes up to the third level of
hierarchy, \emph{i.e.}, only the first four digits of the PACS codes.
This is a good choice for longitudinal analysis: at the third level of
hierarchy, the PACS codes represent the subfields of physics well and all
PACS codes that have been listed in the papers extend at least to this
level. Furthermore, there are more fluctuations in the deeper levels --
the PACS codes change over time, as the classification scheme is regularly
revised by AIP.

\noindent
{\bf Network construction:}
For constructing the networks, we consider the individual PACS codes as
nodes, such that links between them indicate that they have appeared in the
same article. In order to follow the time evolution of this system, we create
yearly aggregated networks by considering all articles published in a
given year. We then extract the largest connected components (LCC) for all the
yearly aggregated PACS networks; all network properties in this paper have been
calculated for LCCs. For all years, the LCC's correspond to almost the whole network ($>99.5\%$).

The weight of the link between the PACS code nodes $i$ and $j$ is defined as 
$w_{ij}=\sum_p\frac{1}{n_p-1}$, where the sum runs over the set of papers in which the
PACS codes $i$ and $j$ appear together, and $n_p$ is the number of PACS codes used in paper $p$. This
 ensures that the strength of each node, $s_i = \sum_j
w_{ij}$, equals the number of articles where the PACS code has been listed~\cite{Newman01a} 
(excluding articles with single PACS codes that are not part of the network). 

\noindent
{\bf Spearman rank correlation, and dissimilarity coefficient:}  If
$r_{1\dots N}^t$ represent the degree (strength) ranks of the PACS codes for year $t$, then
the Spearman rank correlation $C^S$ between the years $t$ and $t'$ is
defined as
\begin{equation}
  C^S_{tt'} = \frac{\sum_i [ r_{i}^{t} - \langle r^{t} \rangle ]
  [r_{i}^{t'} - \langle r^{t'} \rangle ]}
  {\sqrt{\sum_i [r_{i}^{t} - \langle r^{t} \rangle]^2
  \sum_i[r_{i}^{t'} - \langle r^{t'} \rangle]^2}},
  \label{eq:spearman}
\end{equation}
where $\langle \dots \rangle$ represents the average over all nodes. 
From $C^S$ we calculate the dissimilarity coefficient $\zeta \equiv
1-(C^{S})^2$, where $\zeta \in [0,1]$, with low values indicating
that the rank of the individual nodes remain invariant over
time~\cite{Kossinets06}.

\noindent
{\bf Weighted overlap:} In a unweighted network, the overlap is used to
determine the similarity in the neighborhood of two nodes~\cite{Onnela07}.
However, if the network is weighted and the link weight distribution
is heterogeneous, one should put more significance on links having large
weights.
In order to do this we define the weighted version of the neighborhood
overlap $O_{ij}^w$ between nodes $i$ and $j$ as
\begin{equation}
  O_{ij}^{w}=\frac{W_{ij}}{s_i+s_j-2\times w_{ij}-W_{ij}}
  \label{eq:overlap1b}
\end{equation}
where $W_{ij} = \sum_{k\in \Lambda_i \cap \Lambda_j}(w_{ik}+w_{jk})/{2}$
and $\Lambda_i$ denotes the neighborhood of node $i$. Thus, $O_{ij}^{w}=0$
if the two nodes $i$ and $j$ have no common neighbors, and $O_{ij}^{w}=1$
if all of their strength is associated with links to common neighbors
(except for the weight of the link joining $i$ and $j$, if any).

\noindent
{\bf $k$-core analysis:} 
We start by
recursively removing nodes that have a single link until no such nodes
remain in the network. These nodes form the 1-shell of the network
($k^s$-shell index $k^s=1$).
Similarly, by recursively removing all nodes with degree 2, we get the
2-shell. We continue increasing $k$ until all nodes in the network have
been assigned to one of the shells. 
The union of all the shells with index greater than or equal to
$k^s$ is called the $k^s$-core of the network, and the union of all
shells with index smaller or equal to $k^s$ is  the $k^s$-crust
of the network (see also Supplementary Information).


\section*{Acknowledgments}
Financial support from EU's 7th Framework Program's
FET-Open to ICTeCollective project no. 238597 and by the Academy of
Finland, the Finnish Center of Excellence program 2006-2011, project no.
129670 are gratefully acknowledged. We would like to thank S Sanyal
and A Basu for helpful discussions.

\section*{Author Contributions}
All authors designed the research and participated in the writing of the
manuscript. RKP collected the data, analysed the data and  performed the
research.

\section*{Additional Information}
\subsection*{Competing financial interests}
The authors declare no competing financial interests.


\section*{Supplementary Information}
\begin{figure}[t!]
\begin{center}
\includegraphics[width=1.00\linewidth]{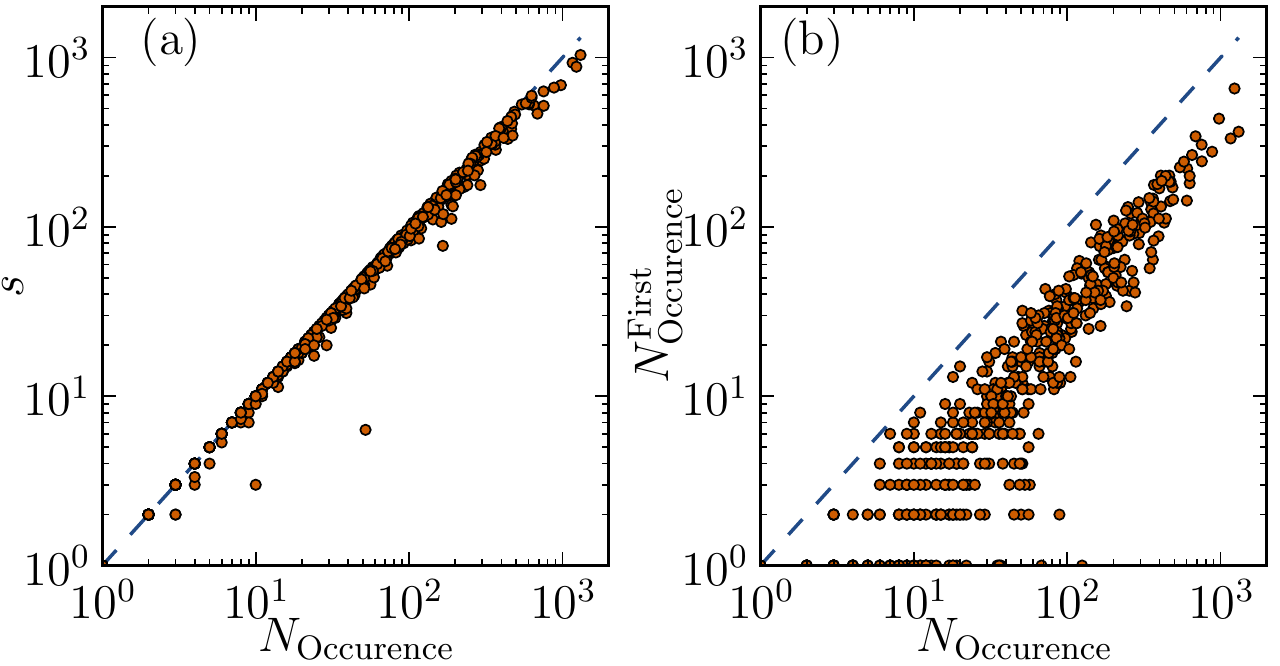}
\end{center}
\caption{The number of times a PACS code has appeared in articles against
(a) the node strength and (b) the number of times it has appeared as the primary code.
The dashed lines show a linear dependence where the quantities
are always equal. The plots are for year
2009, while data for other years indicate qualitative similarity.} 
\label{fig:nPaper}
\end{figure}

\subsection*{Papers with single PACS codes; primary and secondary codes}

For our analysis, we have ignored all papers with a single PACS code. 
Such papers are rather rare, as can been seen by plotting the strengths
of PACS-code nodes in our networks (where single-PACS-code papers are not included)
against the true number of papers where their PACS codes have appeared,
using the entire data set [Fig.~\ref{fig:nPaper}~(a)]. It is evident that
these two quantities are very similar to each other.

It may also be possible that some of the PACS codes frequently appear as
the primary (first) PACS code in an article, and could thus be considered
more important than codes that appear mainly as secondary codes. 
In order to check this, in Figure~\ref{fig:nPaper}~(b) we plot the total number of appearances
of a PACS code against the number of times it has appeared as the primary code.
Although there are some PACS that mainly appear as secondary code, e.g.,
``27.10-Properties of specific nuclei listed by mass ranges $A \leq 5$'',
``02.70-Computational techniques; simulations'', etc., most
of them do appear both as primary as well as secondary code.

\subsection*{Evolution of network properties}
\begin{figure}[h!]
\begin{center}
\includegraphics[width=1.00\linewidth]{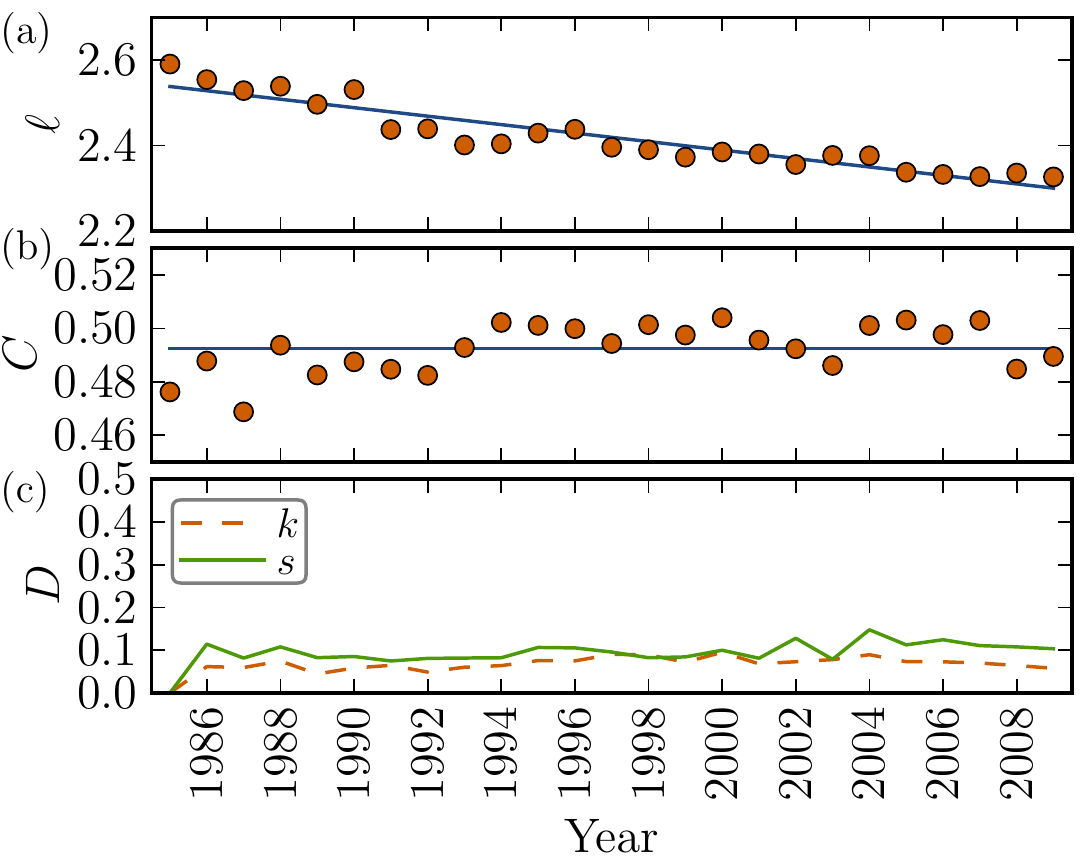}
\end{center}
\caption{Time evolution of
the (a) average path length $\ell$ and (b) the clustering coefficient $C$ of
the PACS network with time. The solid line in (a) indicates a a linear decrease
of $0.01$ in $\ell$. The line in (b) shows that 
clustering fluctuates at values around $C=0.49$ throughout the period of study.
(c) The Kolmogorov-Smirnov statistics, $D$,
comparing the degree and the strength distributions of the year 1985 with the
distributions of subsequent years, indicating stationarity.} 
\label{fig:network_evolution}
\end{figure}

As seen in the main paper, the average degree, $\langle k \rangle$, of PACS networks
increases linearly~[Fig.1~(c) of main paper]. As a result,
the average path length in these networks, $\langle \ell \rangle$,
decreases linearly over this period~[Fig.~\ref{fig:network_evolution}~(a)].
These features indicate that more papers joining different sub-fields of
physics are appearing, leading to an increase of connectivity between them. 
However, the clustering coefficient of the
network turns out to be constant over this
period~[Fig.~\ref{fig:network_evolution}~(b)], suggesting that the local
connectivity of the networks remains almost constant compared to the global
connectivity.

To quantify the similarity between the degree distributions of the PACS networks
of different years,
we measure the Kolmogorov-Smirnov statistics~\cite{Massey51} $D$ of the
degrees of year 1985 with the corresponding distributions of the subsequent
years. Figure~\ref{fig:network_evolution}~(c) indicates that the
distributions do not change much over time, as $D$ remains at a roughly constant, 
low value over this
period. Repeating the above analysis with strength distributions
reveals the same behavior.
  In Table~\ref{tab:ksStats} we show the KS distance as well the
  statistical significance (p values) between the degree (and strength)
  distribution across different years (those shown in Fig.~2 of the main paper).
  If the KS distance is small or the p-value is high, then we cannot
  reject the null hypothesis that the distributions of the two samples are
  the same.  We found that one cannot reject the hypothesis that all
  the degree distribution in Fig. 2~(a) of the main paper are similar to each other
  ($p>0.15$). However, there is more variation across the strength
  distributions. For example, we found that the distribution of 1985 is
  different from 2009 ($p<0.01$). However, for other distributions one cannot
  reject the hypothesis that the distributions are similar ($p>0.05$).

\begin{table}[t!]
  \centering
\begin{tabular}{|l|l|l|l|l|}
    \hline
    Properties & Year & 1993 & 2001 & 2009 \\ 
    \hline
    \multirow{3}{*}{Degree}
    & 1985 & 0.06 (0.36) & 0.07 (0.18) & 0.06 (0.35) \\ 
    & 1993 & & 0.04 (0.63) & 0.06 (0.33) \\
    & 2001 & & & 0.05 (0.31) \\ \hline
    \multirow{3}{*}{Strength}
    & 1985 & 0.08 (0.08) & 0.08 (0.07) & 0.10 (0.01) \\ 
    & 1993 & & 0.07 (0.09) & 0.06 (0.27) \\ 
    & 2001 & & & 0.06 (0.21) \\ \hline
  \end{tabular}
  \caption{
    Kolmogorov-Smirnov (KS) test to assess whether the distributions
    of different years differ significantly.  If the KS distance is small
    or the p-value is high, then we cannot reject the null hypothesis that
    the distributions of the two samples are the same.}
  \label{tab:ksStats}
\end{table}

\begin{figure}[t!]
\begin{center}
    \includegraphics[width=0.99\linewidth]{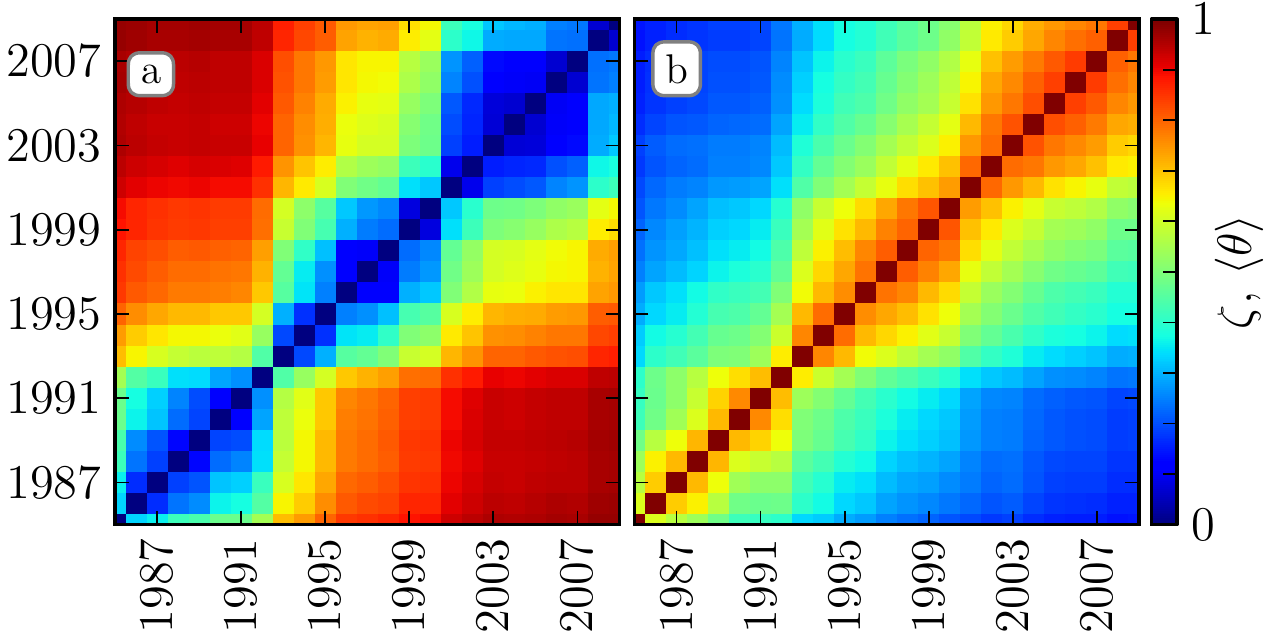}
\end{center}
\caption{
(a) The dissimilarity coefficient $\zeta_{tt'}$ between the node degree ranks of
different years. A small value of $\zeta$ for any two subsequent
years indicates that individual node degree ranks for a given year do
not change much with respect to the corresponding value of immediate
next year. However, this correlation clearly decreases with time. Further, the
blocks in $\zeta_{tt'}$-matrix indicates that the ranks of node degrees
as well as strengths are highly correlated between 1985-1992,
1993-2000 and 2001-2009, whereas between these regions the correlation
is low. The corresponding $\zeta_{tt'}$-matrix for strength ranks of nodes behaves
similarity. (b) Plot of the average Tanimoto coefficient $\langle \theta
\rangle_{tt'}$ which shows that the weighted network structure remains
rather similar for nearby years.}
\label{fig:rankCorrelation}
\end{figure}

Although the shapes of the degree and strength distributions remain same,
the degrees and strengths of the individual nodes do vary in time. 
Fig.~\ref{fig:rankCorrelation}~(a) shows the dissimilarity coefficient matrix $\zeta_{t,t^{'}}$ calculated 
from the rank-correlation
matrix $C_S$ for node degrees between all pairs of years $t$, $t{'}$ (see Methods).
As expected, we observe that the rank order is fairly similar
for consecutive years and this similarity decreases with time. However,
the matrix also shows the presence of a block structure of high
similarity during the periods 1985-1992, 1993-2000 and 2001-2009. The block
structure suggests that the degree ranks of the PACS were
more stable during these periods, and that there were major changes from one
period to another. To find whether only the
characteristics of individual nodes have changed at these points
or whether there are changes in network structure, we consider the similarities 
between local
neighborhoods of nodes for different years. We quantify this with the
Tanimoto coefficient, which is a weighted extension of the Jaccard coefficient, defined as
\begin{equation}
  \theta_{ii}(tt{'})= \frac{\sum_j w_{ij}(t)  w_{ij}(t{'})}
  {\sum_j \left[ w_{ij}^2(t) + w_{ij}^2(t^{'})- w_{ij}(t)
  w_{ij}(t{'})\right] } ,
\end{equation}
where $w_{ij}(t)$ and $w_{ij}(t{'})$ are the weights of the links between
nodes $i$ and $j$ for the years $t$ and $t{'}$, respectively. We then
measure the overall neighborhood similarity of the networks for
different years by considering the weighted average over nodes
\begin{equation}
  \langle \theta \rangle (tt{'}) = \frac{\sum_i [s_{i}(t)+s_{i}(t{'})]\theta_{ii}(tt{'})}
  {\sum_i [s_{i}(t)+s_{i}(t{'})]}
\end{equation}
where $s_{i}(t)$ and $s_{i}(t{'})$ are the strengths of node $i$ for the
years $t$ and $t{'}$, respectively. A high value of $\langle \theta
\rangle$ indicates that the network structure (including link weights)
is relatively invariant. The similarity matrix $\langle \theta \rangle_{tt{'}}$
is shown in Fig.~\ref{fig:rankCorrelation}~(b); again, the networks
of consecutive years appear rather similar.
Further, a
block structure is evident, exhibiting an increased network
similarity for the above periods.

\begin{figure}[ht!]
  \begin{center}
    \includegraphics[width=1.0\linewidth]{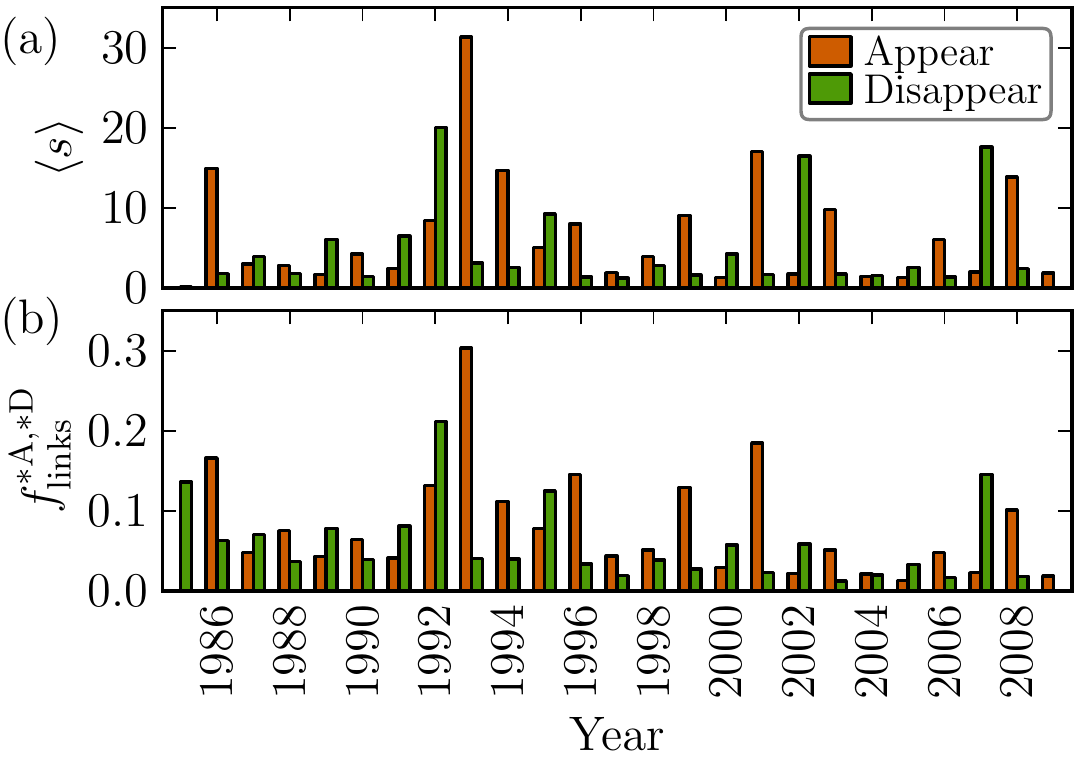}
  \end{center}
  \caption{(a) Average strength of the appearing and the disappearing
  nodes. (b) Fraction of links to and between appearing (disappearing)
  nodes as compared to the total number of appearing (disappearing) links
  in a given year.} 
  \label{fig:nLinksDist}
\end{figure}

To determine the reason behind this observation, we consider the appearing
and disappearing nodes. The average strengths of nodes appearing in
the years 1986, 1993, 2001 and 2008 have been higher compared to other years
[Fig~\ref{fig:nLinksDist}~(b)]. Further, the average strengths of nodes
disappearing in the years 1992, 2002 and 2007 are also relatively high. This
means that many important PACS codes appeared and disappeared during these
years. Next, we focus on the appearing and disappearing links in each
year. We have previously observed in [Fig.1~(d) of main paper] that
roughly same fraction of links appear and disappear every year. 
However, the ratio of links to and between the appearing nodes as compared
to the total number of appearing links in a given year,
$f_{\mathrm{links}}^{\mathrm{*A}} = \sum_{i \in {\mathrm{Appear}}}
k_{i}/N_{\mathrm{links}}^{\mathrm{A}}$ fluctuates with time.  Similarly, ratio
of links to and between the disappearing nodes (just before they disappear)
as compared to the total number of disappearing links in a given year,
$f_{\mathrm{links}}^{\mathrm{*D}} = \sum_{i \in {\mathrm{Disappear}}}
k_{i}/N_{\mathrm{links}}^{\mathrm{D}}$ also varies with time.
Fig~\ref{fig:nLinksDist}~(b) shows that in years, 1986, 1993 and 2001 more
newly appearing links were connected to newly born nodes as compared to the other
years, while in years 1985, 1992 and 2007 more links disappeared due to
nodes disappearing. Thus, there is relatively more change in network
structure during these years due to the high degree of appearing and
disappearing nodes. We found that many of these newly appearing PACS
codes were
introduced to refine the sub-field and thus replace an existing code that
did not represent the field well whereas others were introduced as a result
of discovery of new concepts. 

\subsection*{Comparison of empirical network with null models}
We have compared the structural properties of the PACS network with a
randomized ensemble of networks where the PACS codes are reshuffled among papers. This provides a
null model giving insight into whether the observed properties are
expected by chance as a consequence of the constraints inherent in the system.  In
Table~\ref{tab:randomizedProps1}, we report the number of nodes $N$, the number
of links $L$, clustering coefficient $C$, the average path length $\ell$
and the average weight of the links $w$ of the network for years 1985, 1993,
2001, 2009 and the corresponding randomized versions that are obtained by
randomly shuffling the PACS codes across papers. We first observe that
there are
more links in the randomized network compared to the empirical
network because the 
PACS pairs that appear frequently in the empirical network are now less likely
to be seen together. Instead, each member of the pair are more likely to
appear together with other codes, forming new
links. This leads to an increase in the clustering coefficient,
decrease in the average link weight and also decreases the average path length.

\begin{table*}[t]
  \centering
\begin{tabular}{|l|l|l|l|l|l|l|l|l|l|}
    \hline
    Year &  $N$ & $C$ & $C^{\mathrm{rand}}$ & $L$ &
    $L^{\mathrm{rand}}$ & $\langle w \rangle$ & 
    $\langle w \rangle^{\mathrm{rand}}$ &  $\ell$ &
    $\ell^{\mathrm{rand}}$ \\ \hline
    1985 & 438 & 0.48 & 0.56$\pm$0.009 & 4688 & 9627$\pm$41 & 1.53 &
    0.79$\pm$0.003 & 2.6 & 2.1$\pm$0.009 \\ \hline
    1993 & 503 & 0.49 & 0.65$\pm$0.006 & 7604 & 16193$\pm$56 & 1.89 &
    0.95$\pm$0.003 & 2.4 & 2.0$\pm$0.006 \\ \hline
    2001 & 614 & 0.50 & 0.65$\pm$0.005 & 10797 & 23438$\pm$66 & 1.79 &
    0.89$\pm$0.003 & 2.4 & 2.0$\pm$0.005 \\ \hline
    2009 & 620 & 0.49 & 0.67$\pm$0.005 & 12826 & 29294$\pm$65 & 1.99 & 
    0.95$\pm$0.002 & 2.3 & 1.9$\pm$0.004 \\ \hline
  \end{tabular}
  \caption{
    Comparison of the properties of the real network with the randomized
  null model. The network for the null models are created from an ensemble
where the PACS codes are reshuffled among the papers. The properties of the
random network are averaged over 100 different realizations.}
  \label{tab:randomizedProps1}
\end{table*}

\begin{figure*}[t!]
\begin{center}
\includegraphics[width=0.9\linewidth]{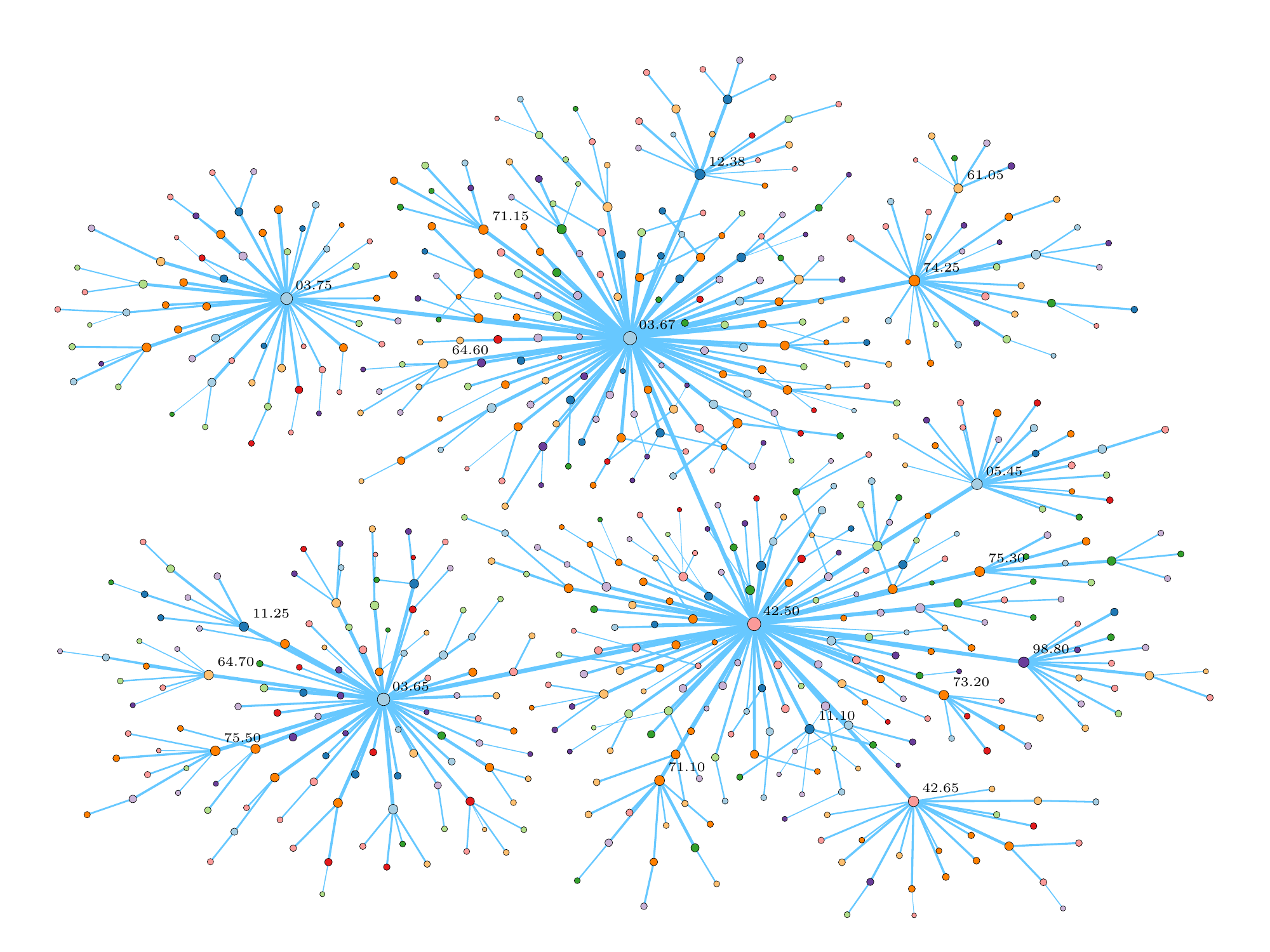}
\end{center}
\caption{The maximum spanning tree of the randomized version of the PACS
network of 2009. Compared to the empirical network, there is a clear lack
of hierarchical structure, as most nodes are connected to a small number of
high-degree nodes. Further, the modularity in terms of frequent connections between 
nodes in similar categories is also lost.}
\label{fig:mst_rand}
\end{figure*}

This randomization process also destroys the hierarchical structure of the
network. In Table~\ref{tab:randomizedProps2}, we show the fraction of links
that connect nodes at different hierarchical distances ($h=0,1,2$). We
compare it with the corresponding fraction in the null model. As
expected,
most of the links now connect nodes that are hierarchically different and
very few links connect nodes that are hierarchically similar. For
example, in
2009, 61\% of links in the empirical network connect nodes that are
dissimilar (h=0), which is much lower that the 87\% of the links that
appear in the null model. Furthermore, 12\% of the links are between similar
nodes (h=2), which is much larger that 2\% of such links in the randomized
network. This loss of hierarchical and modular structure can be seen in the
minimum spanning tree of the randomized version of the PACS network of
2009 [Fig~\ref{fig:mst_rand}]. 

\begin{table}[t]
  \centering
\begin{tabular}{|l|l|l|l|l|l|l|}
    \hline
    Year & $f_{\mathrm{link}}^{h=0}$ &
    $f_{\mathrm{link}}^{h=0}|_{\mathrm{rand}}$ & $f_{\mathrm{link}}^{h=1}$ &
    $f_{\mathrm{link}}^{h=1}|_{\mathrm{rand}}$ & $f_{\mathrm{link}}^{h=2}$ &
    $f_{\mathrm{link}}^{h=2}|_{\mathrm{rand}}$ \\ \hline
    1985 & 0.52 & 0.85$\pm$0.003 & 0.34 & 0.13$\pm$0.002 & 0.14 & 0.02$\pm$0.0012 \\ \hline
    1993 & 0.57 & 0.85$\pm$0.002 & 0.30 & 0.12$\pm$0.002 & 0.12 & 0.02$\pm$0.0008 \\ \hline
    2001 & 0.60 & 0.86$\pm$0.002 & 0.28 & 0.11$\pm$0.001 & 0.13 & 0.02$\pm$0.0006 \\ \hline
    2009 & 0.61 & 0.87$\pm$0.001 & 0.28 & 0.11$\pm$0.001 & 0.12 & 0.02$\pm$0.0005 \\ \hline
  \end{tabular}
  \caption{
    Comparison of the fraction of links between nodes in different
    hierarchical level in the real network with the randomized null model.
    The network for the null models are created from an ensemble where the
    PACS codes are randomly shuffled among the papers. The properties of the
  random network are averaged over 100 different realizations.}
  \label{tab:randomizedProps2}
\end{table}

\subsection*{Microdynamics of new links between dissimilar and similar nodes}
\begin{figure}[t!]
\begin{center}
  \includegraphics[width=1.0\linewidth]{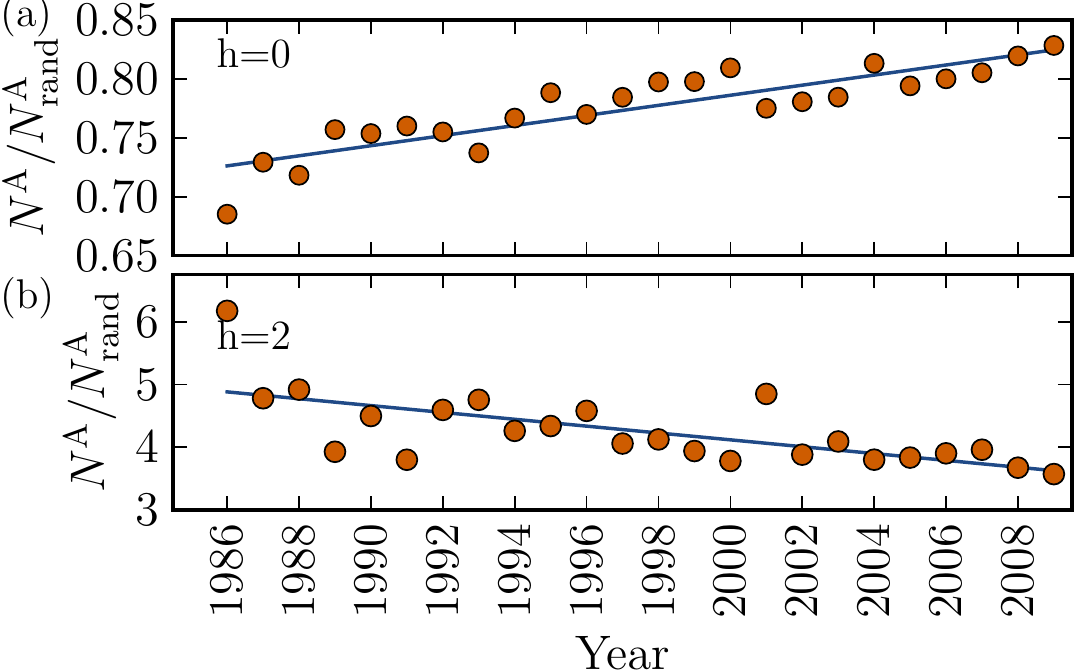}
\end{center}
\caption{The number of newly appearing links $N^A$ falling between PACS codes that are dissimilar $h=0$ or similar
to the 2nd level of the PACS hierarchy ($h=2$), normalized by the expected number $N^A_{\mathrm{rand}}$ in a null model
where all new links are placed randomly in the network.}
\label{fig:newlinks_relative}
\end{figure}

In [Fig.3~of the main paper], we have shown that a substantial and increasing 
fraction of new links connects nodes that belong to different level 1 PACS
categories ($h=0$), whereas the fraction of new links to similar $h=2$
nodes is decreasing with time. However, because of the hierarchical nature
of the PACS tree, there are many more node pairs with $h=0$ than with $h=2$,
and thus even randomly placed links would more often fall between $h=0$ nodes.
Thus, in theory, the increasing number of new links joining $h=0$ nodes might
be explained by the increasing number of PACS codes. In order to verify 
the existence of a real trend, we have plotted the number of new
links between $h=0$ or $h=2$ nodes, $N^A$, normalized by the 
corresponding number $N^A_{\mathrm{rand}}$ in a randomized null model where all the $N^A$ links
are placed randomly. Fig.~\ref{fig:newlinks_relative} shows that the
increasing trend for new links between $h=0$ nodes is present even with this normalization,
and the likelihood of new links connecting dissimilar PACS branches is
thus increasing with time.

\subsection*{Microdynamics of new nodes}
In the main text of the paper we have explained the method to determine  
similar connectivity pattern for disappearing PACS codes. 
We can perform a similar analysis focusing on PACS codes that are newly
introduced. For each of the newly introduced PACS code $i$, we find their
peak years $t^\ast$ with the highest number of published papers and
determine the network neighborhood $\Lambda_{i,t^\ast}$ corresponding to
the peak year. We then calculate the overlap of this neighborhood with the
neighborhoods of all nodes in the network at year $t_i-1$, where $t_i$ is
the year when $i$ appeared. We then choose the node $j$ that has the
maximum overlap with $\Lambda_{i,t^\ast}$, and thus has the most similar
link pattern with $i$ at its peak. As in the case of discontinued PACS, we
found that as the maximum strength of the introduced PACS codes increases,
the maximum overlap also increases. Further, it is seen that only $\sim
10\%$ of new codes appear to replace discontinued codes, and thus the
majority of new codes seem to correspond to emerging new subfields
[Fig.4~(h)~of main paper].

\subsection*{Properties of unstable nodes and links}
\begin{figure}[t!]
\begin{center}
\includegraphics[width=0.98\linewidth]{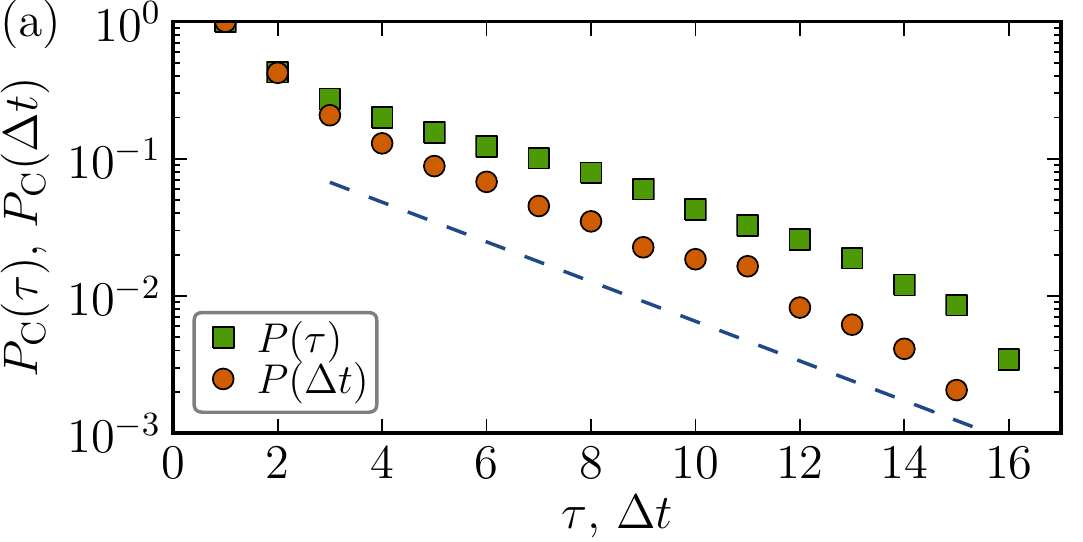}
\includegraphics[width=0.98\linewidth]{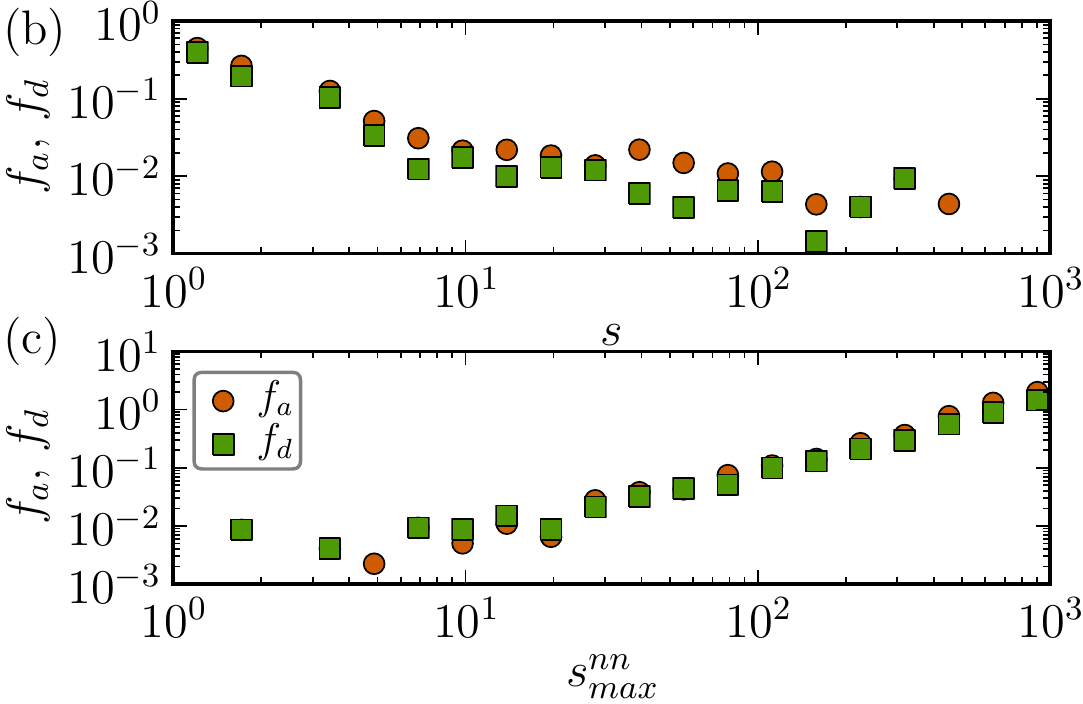}
\end{center}
\caption{Properties of appearing and disappearing nodes. (a) Distribution
of the time of existence $\tau$ and the period of absence $\Delta t$ of the
transient nodes (nodes that both appear and later disappear). The dashed line
indicates an exponential decay, $\sim \exp(-t/3)$. The fraction of appearing
and disappearing nodes of a given (b) strength $s$ and (b) maximum strength
of neighbors $s_{\mathrm{max}}^{\mathrm{nn}}$, as compared to other nodes in the
network.} 
\label{fig:adnodes}
\end{figure}

The PACS network displays turnover in terms of both nodes and links.
Here we focus
on those nodes and links that appear (\emph{i.e.}~are present at year $t$
but not at $t-1$) or disappear \emph{i.e.}~are present at $t$ but not at
$t+1$) during the period of
study (1985-2009). As seen in [Fig.1~(d) of main paper],
the percentage of appearing and disappearing nodes is between 5\%-10\% per year.
We first focus only on \emph{transient} nodes that appear and later disappear
during the observation period. To characterize the transient nodes, we
consider the time for which they are continuously present, $\tau$. 
As transient nodes may reappear in the network after their disappearance, 
we also measure the time of their absence, $\Delta t$. The distributions
of both quantities
decay exponentially, $P(\tau) \propto \exp(-\alpha \tau)$ and $P(\Delta t)
\propto \exp(-\beta \Delta t)$, with exponent $\alpha \sim \beta \sim
1/3$~[Fig.~\ref{fig:adnodes}(a)]. This means that nodes that are present
(absent) for three consecutive years are $1/e$ times less likely to
disappear (appear). 

Next, we compare the properties of all nodes that
appear or disappear during the observation period with other nodes in the network. We define
$f_a(s)$ as the fraction of nodes of strength $s$ that appear during
the period of observation,
\begin{equation}
  f_a(s)=\frac{\sum_t N_t^a(s)}{\sum_t N_t(s)},
\end{equation}
where $N_t(s)$ is the number of nodes with strength $s$ at time $t$ and
$N_t^a(s)$ is the number of nodes with strength $s$ that appear between $t$
and $t+1$. We similarly define $f_d(s)$, the fraction of nodes of strength
$s$ that disappear during the period of observation~\cite{Gautreau09}. Most of these appearing and disappearing nodes have low
strength, indicating they were used in very few papers at the time of
appearance or just before disappearance~[Fig.~\ref{fig:adnodes}~(b)]. However, a few nodes with high
strength appear or disappear with non-negligible
probability. As the degree and the strength of
the nodes are related, the $f_a$ and $f_d$ behave very similarly with the
node degree (not shown). 
When measuring $f_a$ and $f_d$ as a function of the maximum degree of the node's
neighbor, it is seen that appearing and disappearing nodes are mainly
connected to hubs~[Fig.~\ref{fig:adnodes}~(c)]. Thus, most of the appearing
nodes get connected to nodes of high strength and degree and the
neighbors of high-strength and high-degree nodes are more likely to disappear, as
compared to the neighbors of low strength and non-hub nodes.

\begin{figure}[t!]
\begin{center}
\includegraphics[width=0.99\linewidth]{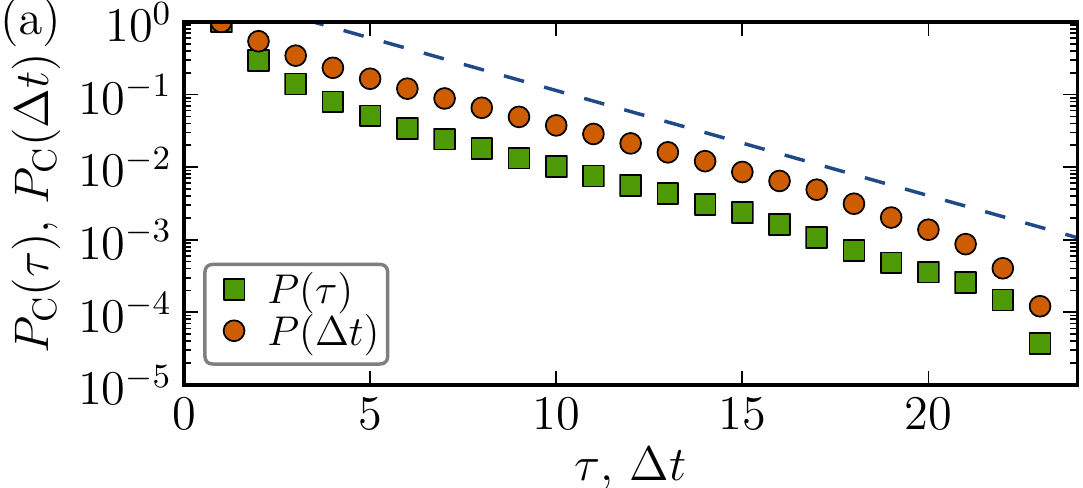}
\includegraphics[width=0.99\linewidth]{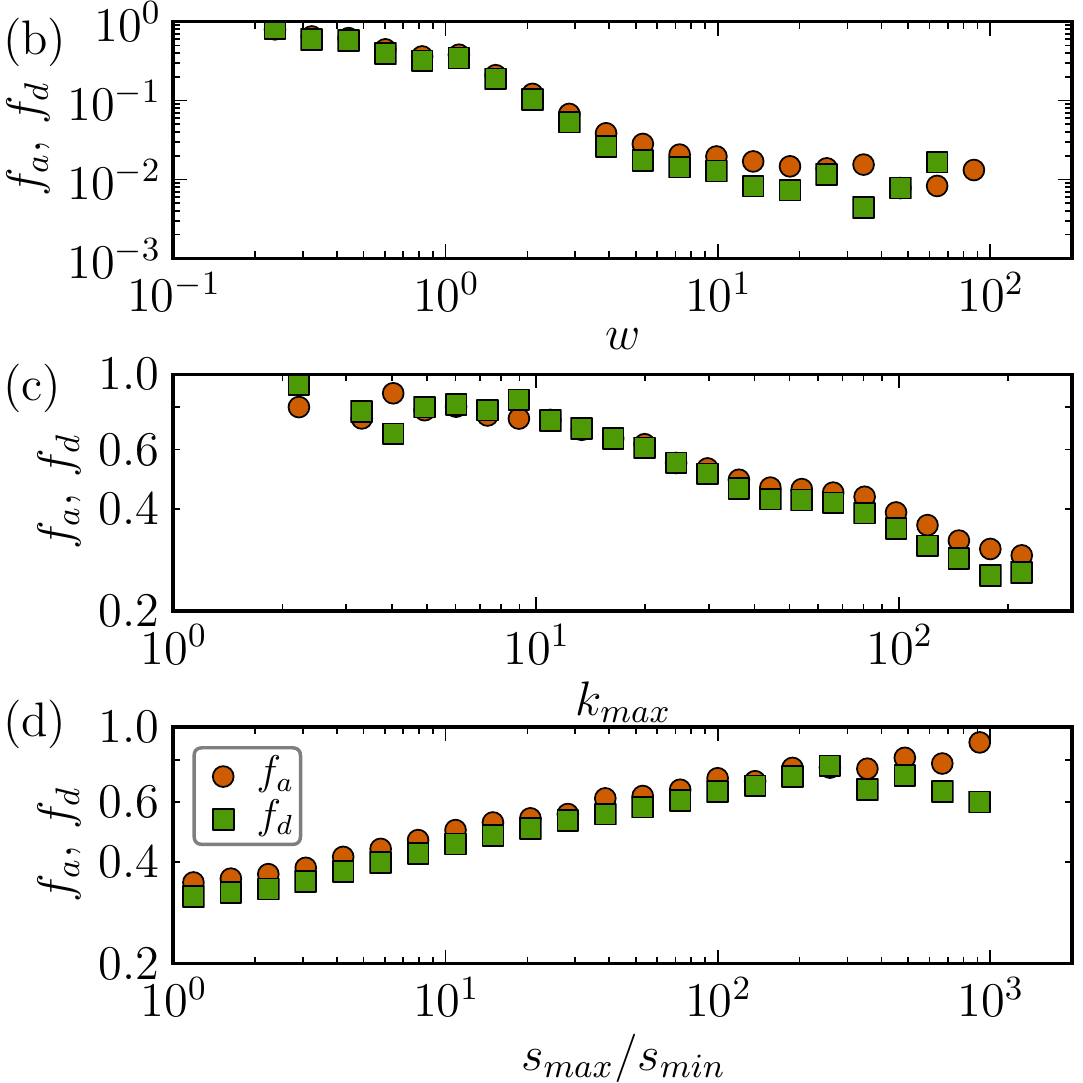}
\end{center}
\caption{Properties of appearing and disappearing links. (a) Distribution
of the time of existence and period of absence for the transient links
(those which appear as well as disappear). The line indicates an
exponential decay, $\exp(-t/3)$. Fraction of appearing and disappearing
links of a given (b) link weight $w$, (c) maximum degree of the connecting
nodes $k_{\mathrm{max}}$ and (d) ratio of strength of nodes connected by
the link, $s_{\mathrm{max}}/s_{\mathrm{min}}$, as compared to the overall
links in the network.} 
\label{fig:adlinks1}
\end{figure}

Next, we focus on the links that appear or disappear during our period of
observation; as seen in [Fig.1~(d) of main paper], about 40
percent of links appear and a similar number of them disappear
every year. We first consider only the transient links that
appear and then disappear during the period of observation. In
Fig.~\ref{fig:adlinks1}~(a)
we show the distribution for the period for which they were present
continuously, $\tau$, and the period of absence, $\Delta t$,
defined as for the nodes. Again, both
distributions decay exponentially with an exponent of $-1/3$, similarly to
the behavior observed for transient nodes. This suggest that most of these
links appear or disappear as new nodes are introduced to the network or nodes
leave the network, respectively.  This behavior is different from the node
and link dynamics of air transportation network~\cite{Gautreau09} where the nodes are mostly
stable and the distribution of link's absence and presence decays as a
power law. This means that in the PACS network
links that are absent for a long time are much less likely to reappear, and
links that are present for a considerable period are much less likely
to disappear, as compared to the case in the airport network. This may be
related to the economic constraints operating in the airport network that
make commercially unenviable links more likely to disappear and the
profitable links more likely to appear.

As we did for nodes, we also compare the properties of all appearing and
disappearing links with overall properties of links.The fraction of
links of weight $w$ that disappear during the time-period 1985-2009 is
defined as 
\begin{equation}
  f_d(w)=\frac{\sum_t N_t^d(w)}{\sum_t N_t(w)},
\end{equation}
where $N_t(w)$ is the number of links with weight $w$ at time $t$ and
$N_t^d(w)$ is the number of links with weight $w$ that disappear between
$t$ and $t+1$. The fraction $f_a(w)$ of links of weight $w$ that
appear is defined as for the nodes. Most of the appearing and
disappearing links have low weight; however, links with high weight may
also appear and disappear with a non-negligible
probability~[Fig.~\ref{fig:adlinks1}~(b)]. We also measure $f_a$ and $f_d$
as a function of the maximum degree of the two nodes that the link connects.
We find that the most of the links which appear or disappear are between
non-hubs~[Fig.~\ref{fig:adlinks1}~(c)]. Similarly, we measure $f_a$ and
$f_d$ as a function of the ratio of the $s_{\mathrm{max}}/s_{\mathrm{min}}$ of a link, where $s_{\mathrm{max}}$
and $s_{\mathrm{min}}$ are the maximum and the minimum strength of the
nodes joined by the link. Fig.~\ref{fig:adlinks1}~(d) shows that these
links mostly connect nodes of heterogeneous strength.

\subsection*{$k^{s}$-core decomposition and $k$-crust connectivity}
\begin{figure}[b!]
\begin{center}
  \includegraphics[width=1.0\linewidth]{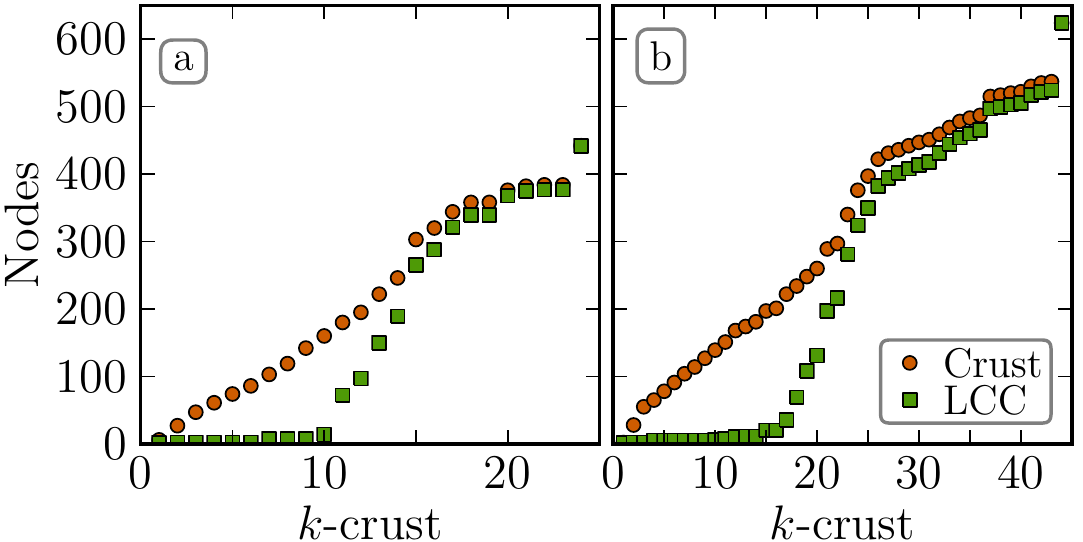}
\end{center}
\caption{The number of nodes and size of the largest connected component in
each of the $k^{s}$-crust for the years (a) 1987 and (b) 2007.} 
\label{fig:kCrust}
\end{figure}

In Figure~\ref{fig:kCrust}, we show the number of nodes and the size of 
the largest connected component (LCC) as a function of the $k$-crust from the $k^s$-core
decomposition of the network of years 1987 and 2007. As expected, both the
number of nodes and the LCC size increase with $k$-crust. For
smaller $k$, the LCC and the crust sizes are different, whereas for
larger $k$ the LCC becomes almost of the same size as the crust. This
feature is different from some other empirically observed
networks~\cite{Carmi07}, where the nucleus plays a crucial role in the
connectivity of the network. In most of these empirical systems, the network
is in general fragmented into multiple disconnected components before the
introduction of the nucleus. However, in the PACS network
the crust is already almost connected even before the
introduction of the nucleus. Therefore, in the PACS network, the nucleus
plays a less important role; e.g., were any dynamical process of information
flow to take place on the network, it would not necessarily need to pass
through the nucleus.

\subsection*{Evolution of publication volumes of PACS codes}
\begin{figure*}[t!]
\begin{center}
\includegraphics[width=0.99\linewidth]{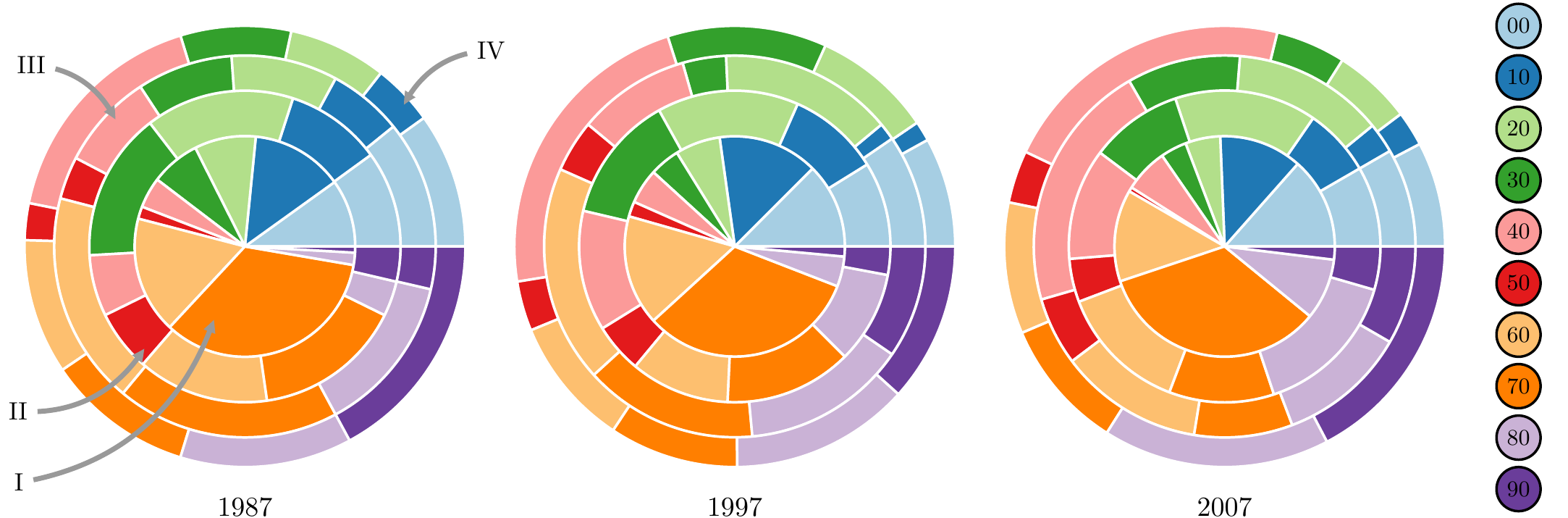}
\end{center}
\caption{Multi-level pie chart for years 1987, 1997 and 2007 showing the time
evolution of the publication volumes of PACS codes which are categorized
according to their field they represent.} 
\label{fig:strengthEvolution}
\end{figure*}
Instead of the $k$-core decomposition and the core indices of PACS codes,
one could argue that the importance of a PACS code might be represented
simply by the number of papers published with it. To compare with the 
$k^s$-shell analysis and the evolution of the core indices of different
codes, as done in [Fig.8~of main paper], we plot a similar 
multi-level pie chart where the regions correspond to the numbers of 
papers with given PACS codes. Again, Region I contains the top 25\%
PACS codes, this time in terms of total publication volume, and Regions II, III,
and IV PACS codes with increasingly lower publication volumes.
As before
we categorize the PACS codes in each of these region with the first
digit of their hierarchy. Although the number of papers for a code and its
$k^s$-shell index are related, Fig.~\ref{fig:strengthEvolution} is very
different from [Fig.8~of main paper]. For each year, all 
fields are represented in each of the four regions. This means that
for all PACS categories, there are sub-categories with high publication volumes
and sub-categories with low volumes. Even the subfields of ``10-The Physics
of Elementary Particles and Fields'' and ``20-Nuclear Physics'' are present in
the region I, whereas
they appear only in the mid and peripheral shells when categorized
according to their $k^s$-shell index. There are no clear trends,
although there is small increase in the number of high-volume
 ``80-Interdisciplinary Physics and Related Areas of Science
and Technology'' PACS codes.

\end{document}